\documentclass[graybox]{svmult}
\usepackage{color}

\usepackage{mathptmx}       
\usepackage{helvet}         
\usepackage{courier}        
\usepackage{type1cm}        
\usepackage[margin=1in]{geometry}
\usepackage{float}
\newfloat{algorithm}{t}{lop}

\usepackage{makeidx}         
\usepackage{graphicx}        
\graphicspath{{Figures/}}
\usepackage{multicol}        
\usepackage[bottom]{footmisc}
\usepackage{algorithm, algpseudocode} 
\usepackage{subcaption}      
\usepackage{amsmath}         
\usepackage{listings}

\makeindex             

\begin{document}
 \titlerunning{ScientoBASE}
 \authorrunning{Gouri Ginde et al.}
  \title*{ScientoBASE: A Framework and Model for Computing Scholastic Indicators of non-local influence of Journals via Native Data Acquisition algorithms}
\author{Author names blinded for review.}
\author{Gouri Ginde, Snehanshu Saha, Archana Mathur, Sukrit Venkatagiri, Sujith Vadakkepat, Anand Narasimhamurthy, B. S. Daya Sagar\vspace{-2ex}\vspace{-2ex}}


\institute{Gouri Ginde \at Department of Computer Science and
        Engineering, PESIT South Campus Bangalore, India \email{gouri.ginde@gmail.com}
\and Snehanshu Saha \at Department of Computer Science and
        Engineering, PESIT South Campus Bangalore, India \email{snehanshusaha@pes.edu}
 \and Sukrit Venkatagiri \at  Department of Computer Science and
        Engineering, PESIT South Campus Bangalore, India \email{95sukrit@gmail.com}}

        \vspace{-2ex}

%
%
\maketitle

\abstract{Defining and measuring internationality as a function of influence diffusion of scientific journals is an open problem. There exists no metric to rank journals based on the extent or scale of internationality. Measuring internationality is qualitative, vague, open to interpretation and is limited by vested interests. With the tremendous increase in the number of journals in various fields and the unflinching desire of academics across the globe to publish in "international" journals, it has become an absolute necessity to evaluate, rank and categorize journals based on internationality. Authors, in the current work have defined internationality as a measure of influence that transcends across geographic boundaries. There are concerns raised by the authors about unethical practices reflected in the process of journal publication whereby scholarly influence of a select few are artificially boosted, primarily by resorting to editorial manoeuvres. To counter the impact of such tactics, authors have come up with a new method that defines and measures  internationality by eliminating such local effects when computing the influence of journals. A new metric, \textbf{Non-Local Influence Quotient} (NLIQ) is proposed as one such parameter for internationality computation along with another novel metric, \textbf{Other-Citation Quotient} as the complement of the ratio of self-citation and total citation. In addition, SNIP and International Collaboration Ratio are used as two other parameters. As these journal parameters are not readily available in one place, algorithms to scrape these metrics are written and documented as a part of the current manuscript. Cobb-Douglas production function is utilized as a model to compute JIMI (Journal Internationality Modeling Index). Current work elucidates the metric acquisition algorithms while delivering arguments in favor of the suitability of the proposed model. Acquired data is corroborated by different supervised learning techniques.  As part of future work, the authors present a bigger picture, RAGIS- Reputation And Global Influence Score, that will be computed to facilitate the formation of clusters of journals of high, moderate and low internationality.}
\vspace{2mm}

\section{Introduction}
\label{sec:1}
In recent times, various authors and research scholars have been exploring means to find suitable and reputed journals for publication of their research work. The drive behind this is to own appreciation or award for the quality work that they do. Also, institutional
assessment and evaluation depends heavily on peer-reviewed publications whether it be for academia or research labs. Generally, the trend observed among various faculties is to publish their research in journals with an 'international' tag attached to it. Thus, evaluating internationality is an open problem owing to the fact that such journals are vast in number; every such entity claims "internationality" but citation and influence are a bare minimum.\\

\par Data collected from IEEE Xplore in the year 2009 showed an exponential increase of 25\% in journal publications, when compared with previous years. A study
conducted by Buchandiran \cite{1} reveals an enormous increase in publication of journals between the years 2004 and 2009, whereby in the year 2009, 6,132 Indian institutions have contributed 23,745 papers out of which 15,880 were from academic institutions. This clearly shows that academic institutions contribute to the majority of such published works. Leonard Heilig and Stefan VoB 
cite{2} observed a significant increase in the number of research publications (only in the area of cloud computing) from 2008 onwards. Elsevier's Scopus covered 15,376 publications till 2014 and Thomson Reuters Web of Science covered 8,262 publications in the same field.\\
 
 \par On the flip side, there exists scam open access publishers who unethically and unprofessionally exploit the open access publishing model for financial rewards. They charge authors for publication fees and publish their work without providing true editorial services as well as other types of services associated with any legitimate journal. This shady publishing practice was first noticed by Jeffrey Beall,an academic librarian and a researcher at the University of Colorado in Denver. He scrutinized and investigated further and based on his findings, published his first compiled list of predatory publishers in 2010 \cite{19}. Continuing on the same line, Jeffrey Beall,\cite{20} regularly updated this list of fake publishers and had put forth certain criteria for categorizing such publications in order to prevent newer scholars from falling prey to such practices.\\
 
 \par The phenomenon of predatory publishing (also known as the dark side of open access publishing) has exploded in recent years with the number of such publications expanding from 53,000 in 2010 to 420,000 in 2014. Accepting articles quickly without peer-review, pursuing academicians to submit articles or to serve on editorial boards, notifying authors of article processing fees only after paper acceptance, improper usage of ISSN and counterfeit Impact Factor values are some of the key indicators which have emerged from the observed working pattern of fake, predatory publishers. Till date, no fool-proof method has been devised to distinguish legitimate publishers from illegitimate ones. \\
\par An abundance of work has been done to evaluate the influence or prestige of scholarly articles and journals. Citation Index, a concept defined by Eugene Garfield (founder of Science Citation Index, SCI and the Institute for Scientific Information, ISI) serves as a link between similar scientific journals and literature. Citation pattern and citation frequency used by Garfield in his foundational work for Web of Science (Thompson Reuters Web of Science) initiated a vast spectrum of research and provided fully indexed and searchable research content. Thompson Reuters then initiated publishing Journal Citation Reports (JCR) to evaluate citation frequency of journals and down-the-line, Impact Factor. \\
\par Another initiative, Elsevier’s Scopus has a vast collection of peer reviewed scholarly journals and citations in scientific, medical, technical and social science domain. Scopus utilizes its database to provide another type of journal metric used for ranking for its journals through the SCImago Journal and Country Rank (SJR) portal \cite{8}. The SJR rank is a score evaluated from the past five years' data addressing a small number of journals. It is claimed that SCI, Thompson Reuters is a little more selective than Scopus.
 The concept of citation index, Impact Factor and SJR ranking provide a limited respite to the above mentioned challenges of distinguishing and ranking legitimate publishers from the fake entities. This gives plenty of motivation and reason to work on proving a journal's credibility and integrity as well as ascertaining the quality, impact and influence of the publications. \\
 
 \par Our initiative, ScientoBASE epitomizes a new approach for evaluating journals in a "height-weight" manner. The database, when complete, will help identify and bring adequate attention to quality journals, including industry practitioner domains, which otherwise would not be possible because certain journals namely Software Quality Professional (SQP) refuse to be indexed. \\
 
 \par The remainder of the paper is organized as follows. Section 2 contains literature survey carried out on existing work related to measuring a journal's internationality vis-a-vis non-local influence. The section also brings forth systemic lags in establishing an unbiased score for ranking of journals. Section 3 defines internationality as perceived by the authors. It presents a schematic view of the approaches used to model internationality. Section 4 presents the algorithmic overflow for calculating internationality. Section 5 discusses in detail different techniques and  algorithms used to collect scholastic parameters for the model described in section 6. These parameters are programatically scraped from multiple web sources such as Google Scholar, IEEEXplore, SCImago and Aminer \cite{25,26}. Once parametric data is acquired, these are then fed into the Cobb Douglas production function; an econometric model that is described in detail in section 6. Section 7 sheds some light on the merits of Source-Normalized Impact per Paper (SNIP) and shows why SNIP - and not Impact Factor - is a good albeit incomplete indicator for estimating non-local influence. Further, the section introduces new metrics, Non-Local Influence Quotient (NLIQ) and Other Citation Quotient and argues in favor of the usefulness of such metrics towards computing internationality. The paper concludes with a discussion on future work embodying ranking and clustering of journals according to internationality in their respective subject areas. The future work is commensurate with the current framework and model proposed in this paper.

\par

\section{Literature Survey}
\label{sec:2}
  Neelam Jangid Snehanshu Saha, Siddhant Gupta, Mukunda Rao J \cite{6,7} in their work used a lightweight approach and introduced a new metric, Journal Influence Score (JIS), which is calculated by applying principal component analysis (PCA) and multiple linear regression (MLR) on citation parameters, extracted and processed from various scholarly articles in different domains, to obtain a score that gauges a journal’s impact. The higher the score, the more the journal is valued and accepted. Journal’s ranking results are compared with ranks of SJR, which internally uses Google's PageRank algorithm to calculate ranks. The results showed minimal error and the model performed reasonably well. Seyyed Mehdi et al. \cite{9} studied the scientific output of fifty countries in the past 12 years. In order to measure the 'quality' and 'quantity' of research output, a two-dimensional map is constructed and analyzed. Clusters are generated after analysis to represent  country wise research output. There exists no ranking mechanism to rank countries with the maximum output in terms of quality and quantity of journals themselves. 
  \par Anup Kumar Das, Sanjaya Mishra \cite{10} discussed how research communities are preferring article-level metrics (ALM) over Journal Impact Factor (JIF) to assess the performance of individual scientists and their  contributions. Gunther K. H. Zupanc \cite{11} also stressed on the unsuitability of using Journal Impact Factor to compare the influence of journals, especially when journals are from different areas. He claims that authors are tempted to publish their work in high-Impact Factor journals instead of journals that are best suited for their research work. A. Abrizah et al. \cite{16} compared the coverage, ranking, impact and subject
categorization of Library and Information Science journals, in which 79 titles were from Web of Science and 128 from Scopus. The prestige factor score of journals from JCR (Journal Citations Report 2010) and SJR (SCImago Journal Rank 2010) was extracted and the difference in ranks was noted. They observed a high degree of similarity in impact factor of titles in both Web of Science and Scopus. At the same time, authors also observed that the two databases differ in the number of journals covered.\\
\par Henk F. Moed \cite{29} introduced a different indicator of journal citation.
impact, Source Normalized Impact per Paper (SNIP). SNIP is defined as the ratio of the journal’s citation count per paper and the citation potential in its subject field. It aims to allow direct comparison of sources in different subject fields. There is no single ‘perfect’ indicator of journal performance. Delimitation of a journal’s subject field does not depend upon some predefined categorization of journals into subject categories but is entirely based on citation relationships. It is carried out on a paper-by-paper basis, rather than on a journal-by-journal basis. SNIP is based on citations from peer-reviewed papers to other peer-reviewed papers. Ludo Waltman et al. \cite{12} have discussed a number of modifications that were recently made to the SNIP indicator. The SNIP indicator considers a source normalized approach to correct the differences in citation practices between scientific fields. The key benefit of this approach is that it does not require the classification of subject fields, where the boundaries of fields are defined explicitly. There are some arguments around the original SNIP indicator’s properties that may be considered counter-intuitive. For instance, it is possible that additional citation has a negative correlation with journal’s SNIP value. \\
\par Gaby Haddow, Paul Genoni \cite{13} defined a new model - Excellence for Research for Australia (ERA) to determine the efficacy of citations measures in order to determine the quality of Australian social science journals. Chiang Kao \cite{33} investigated the contribution of different countries to international repositories of research in industrial engineering journals. After compiling journal data from ISI from 1996 to 2005, it was evident that the USA, UK and China are the top three countries to contribute articles to IE journals and six Asian countries are in the top ten. Yu Lipinga et al. \cite{34} classified common journal evaluation indicators into three categories, namely three first-level indicators. They are, respectively, the indicators on journal impact, on timeliness, and on journal characteristics. The three categories of indicators are correlated with one another, so a structural equation may be established. Then authors calculated the value of three first-level indicators and gave subjective weights to these indicators. This approach provides a new perspective for scientific and technological evaluation, in a general sense.
There are some limitations of this approach:
\begin{itemize}
  \item The availability of basic data and the rationality of modeling bear much upon the evaluation results.
  \item    If there are too many indicators in a scientific and technological evaluation, data availability will be relatively difficult, and the evaluation cost will increase. If indicators are too few, they cannot provide adequate information.
  \item     If the data is inaccurate or wrong, no satisfactory results will be
obtained. In scientific and technological evaluation, sometimes certain data is very difficult to gather.
\end{itemize} 
 \par Gualberto Buela-Casal et al. \cite{14} performed a survey on existing measures of internationality and observed that a valid and quantitative internationality index should differentiate between “quality” and “internationality”. They suggested that in order to measure internationality, suitable weights should be assigned to certain identified parameters using a large-scale census of journal data. They proposed a neuro-fuzzy system to construct an unambiguous journal internationality index. \\

\par Chia-Lin Changa et al. \cite{15} examined the issue of coercive journal 
citations and the practical usefulness of two recent journal performance metrics i.e. Eigenfactor Score, which may be interpreted as measuring “journal influence”, and the Article Influence Score, using Thomson Reuters Web of Science. Authors compared the two new bibliometric measures with the existing ISI metrics, total citations and the 5-Year Impact Factor (5Y-IF) of a journal. It is shown that the sciences and social sciences are different in terms of the strength of the relationship of journal performance metrics, although the actual relationships are very similar. Authors concluded that the Eigenfactor Score (measuring journal influence) and Article Influence  performance metrics for journal are shown to be closely related empirically to the two existing ISI metrics, and hence add little in practical usefulness to what is already known, except for eliminating pressure arising from coercive journal self-citations. \\

\par Predatory publishing has earned a lot of attention (in terms of approval as well as criticism) from different sections of research communities across the globe \cite{21}. Beall's list of predatory journals has been welcomed by many open access supporters, whereas others have raised serious doubts about it's credibility. Walt Crawford \cite{24} in 2014 thoroughly investigated the list and called it a "One Man's List". He concluded that it should be ignored and suggested some steps to evaluate a journal's trustworthiness prior to submission.\\
 \par Step 1: To make a pertinent decision whether "The International Journal of A" is a good target, one must look it up in the Directory of Open Access Journals (doaj.org). If the journal is not in the directory, look for another journal in a similar subject category.\\
\par Step 2: If the journal is in DOAJ, explore its site, its APC policy, quality of English used, its editorial board members - whether they are real people. Otherwise start from step 1.\\
\par Step 3: Check whether article title over the past issues makes sense within the journal's scope or if any author show up repeatedly within the past few issues. If so, go to step 1 again.\\

  \par One can escape from predatory journals utilizing this technique. Nonetheless, it needs a lot of involvement in knowing how to assess journals as there is no scientific model which will do so for us.   
   Additionally, this algorithm is, to a greater extent, a manual investigation and hence ungainly and elaborate.
   Therefore, there is a pressing need to build a complete, end to end web interface that also serves as repository and information visualization toolkit for scientometric evaluation, modeling and analysis. ScientoBASE is designed to serve this purpose and cater to internationality modeling and interface estimation of peer-reviewed journals in the fields of science and technology.

 \section{Definition, Objective and Schematic View}
\label{sec:3}
This section defines internationality and presents an overview of the steps to achieve the end results. The authors would like to take this opportunity to stress that the "internationality of a journal" is defined here as a measure of influence beyond restricted boundaries. These boundaries may be geographical or regional or even cliques or networks of journals. It was observed during the course of this research that citations occur mostly within the journal from which the original citing article was published. The authors believe that in a community which is essentially international by nature, such trends don't bode well. A new metric which is an offspring of this realization, will be elaborated in due course. In order to remain clear about our objectives and dispel any confusion, we reiterate that internationality as defined and measured throughout this paper is a reflection of "non-local influence" and therefore does concur with the standard definition. The basic steps taken to achieve the end results are as follows:

\begin{itemize}
  \item Defining and measuring internationality
  \item Creating a suitable model
  \item Validating the model
  \item Generating granular clusters of “international” journals and conferences (part of future work)
  \item Model the diffusion of internationality (part of future work)  
\end{itemize}

\textbf{Definition}: Internationality of a journal, as proposed by the authors, is a holistic parameterization, of the international aspects of a journal's dimensions. These dimensions include - but are not limited to - quality of publications and measure international span of subscribing readers, authors and reviewers. These additionally evaluate the geographic source of a journal’s citations and the impact it spreads across nations. Authors explore these dimensions in succession and refer to International Collaboration Ratio; a parameter that indicates the ratio of articles whose author affiliations are from distinct nation, has the potential to be a suitable candidate for evaluating a journal's prestige. Likewise, extensive self-citation is a self promoting strategy which is unfairly used by authors to artificially boost their scientific influence, and thus indirectly inflate the publishing journal's impact factor. If used skillfully, this self-citation parameter can provide a good insight to judge a journal's credibility. It must be stated at this juncture, that some of the most common attributes of internationality such as ISSN number, constitution of editorial boards, country of publication and reputation of publishers as input factors are not considered. This is precisely because the authors do not view these attributes as entirely sufficient measures for internationality. Rather, use of such attributes as yardsticks in judging internationality is viewed as impediments towards objectively classifying journals. In recent times most journals have become international by structure (composition od editorial boards, reviewers, submitting authors etc) therefore in an essentially internatnational community of scholastic publication, \textbf{The authors define internationality as assimilation and evaluation of parameters that are \lq truely scholastically international\rq by including significant factors beyond the local manipulation of authors/editors.} 
\\ 
\par There are well-accepted influence measurement parameters used by various web portals. Source-Normalized Impact per Paper (SNIP), allows comparison of sources across and within the same subject field by calculating their citation potential and normalizing their citation impact by dividing their RIP's (Raw Impact per Paper) with the calculated database citation potential. Integrity of academic publications would be at risk if editors coerce authors to cite their journals for enhancing their impact factor. With the intention to weaken the effects of this strategy, authors have introduced a new metric Non-Local Influence Quotient (NLIQ) which is a ratio of "non-local" citations of an article to the total number of citations. Larger the value of NLIQ, more "international" a journal is.  \\

\par Taking all these factors into account, authors proposes a high-level design and methodology to model internationality index that would scrape and assemble the above mentioned parameters and generate journal clusters of high, moderate and low internationality. As already indicated, after computing "internationality", generating granular clusters of journals is a future plan of action. The current work embodies various algorithms for  parameter acquisition and discusses the suitability of the new metric for influence calculation.\\  
\par Empowered by data acquisition techniques, two approaches are put forth for modeling (Fig. 1). The first approach takes data from the Scopus and SJR portals and calculates a journal's score (JIS) \cite{6} using a  multiple linear regression model on the scraped scientific indicators. Second approach, uses non-indexed, non-Scopus/non-Web of Science databases to acquire scientific parameters and evaluate a journal's internationality score generated from a Modeling Index (JIMI, Journal Internationalty Modeling Index) \cite{30, 31}. The approach uses Cobb-Douglas \cite{35,36} and Log Production Model on the parameters scraped from web. The algorithms and procedures are described in section 5 and 6.\\

\par The prestige/internationality of a journal is a convex combination of JIS [ please refer additional files on GitHub, \textbf{32} ] and Internationality Score, represented as-    
Internationality of a journal, 
 \begin{center}
 YI  = ${\alpha}$ JIS + (1 - ${\alpha}$) JIMI ;
\\ $0 < {\alpha} < 1$
 \end{center}
where YI  refers to the internationality score as response
variable(to be sorted in decreasing order), JIS is the influence score obtained from metric JIS, JIMI is the score evaluated from work done using two parameters (JIMI) and ${\alpha}$ is a weight deduced from the cross correlation.
\begin{figure}[h]
\centering
\includegraphics[width=11cm,scale=0.8]{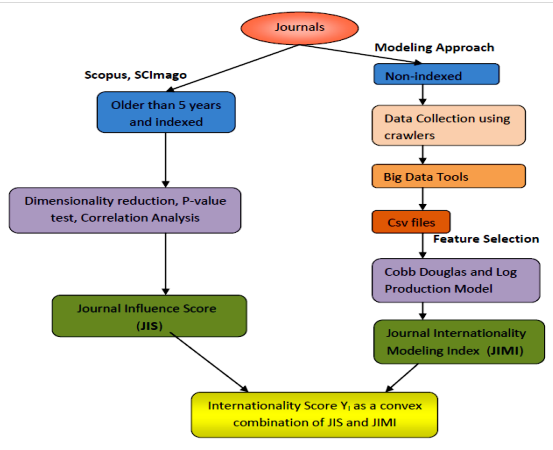}
\caption{Block diagram of Methodology.}
\end{figure}
 For JIS, refer \cite{17} and Appendix I in\cite{32}.
   
 \subsection{Journal Internationality Modeling Index(JIMI)}

 The prestige of an academic journal is derived from quantifiable as well as non-quantifiable factors. Some commonly accepted factors that reflect a journal’s prestige are Impact Factor (IF), Eigenfactor, SCImago Journal Rank (SJR), Source-Normalized Impact per Paper (SNIP),  Impact per Publication (IPP), internationality etc. Impact Factor, as per Thomson Reuter's definition \cite{22} is a measure of the frequency with which an article of a journal has been cited in a particular duration.  The IPP (Impact per Publication) measures the ratio of citations in a year to scholarly papers published in the three previous years divided by the number of scholarly papers published in those same years. When normalized for the citations in the subject field, the Impact per Publication becomes the Source-Normalized Impact per Paper (SNIP). The SJR or SCImago Journal Rank is a measure of the scientific prestige of scholarly sources. SJR assigns relative scores to all of the sources in a citation network.\\
     \par In this section, authors discuss a technique \cite{30} to quantify internationality by
     exploiting a mathematical model, which determines the internationality of a journal by using two major metrics - Source-Normalized Impact per Paper (SNIP) and International Collaboration. Here author stresses on the efficacy of such a model and confirm the model theoretically. We prove that the model has a global maxima where a particular value of the inputs (SNIP and International Collaboration) would ensure some maximum value of internationality, subject to a constraint or set of constraints.
     \begin{figure}
      \centering
      \includegraphics[width=11cm]{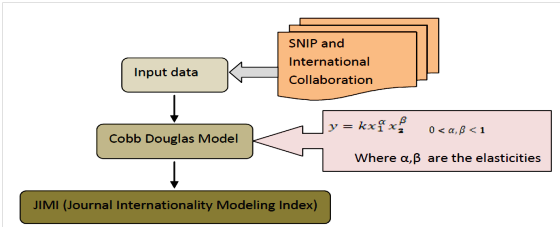}
      \caption{Computation model of Journal Internationality Modelling Index (JIMI).}
      \end{figure}
      
      As shown in Fig. 2, the ,modeling approach uses web scraping technique to extract required features of  various journals to generate CSV data files.
     All these features from various data sources are processed further for only desired features of a journal which will contribute to the evaluation of internationality index. These features are specifically, 
       \begin{itemize}
   
         \item International Collaboration Ratio 
         \item Source-Normalized Impact per Paper (SNIP)
         
         \item Other-Citation Quotient
         \item Non-Local Influence Quotient (NLIQ)
       \end{itemize}
     Using the above features as input parameters to Cobb-Douglas production function \cite{35,36}, the authors intend to measure the internationality index, proposing a score to gauge the influence of peer-reviewed journals.
     \\
     \\ \par \textbf{Sample Data}: This modeling approach uses a consolidated database generated by crawling the web (using software tools) to gather all recent (non-indexed) journals that are older than 3 years and younger than 5 years from Google Scholar. Reasons for selecting this are:
     \begin{enumerate}
\item 	Elsevier considers a 3-year window for SNIP mainly due to the difference in the rates at which subject fields mature, whereas Thomson Reuters has a 2-year and 5-year window for Impact Factor (IF). As noted in section 7.1, one notable advantage of SNIP over IF is that SNIP’s 3-year citation window allows fields that move at a slower pace to be compared with those that advance fairly rapidly, in as fair a manner as possible. Whereas the 2-year IF and 5-year IF only favor one or the other. 
Thus, authors have taken a window of 3-5 years in order to cater to journals in both categories.
	
\item Scraping the data which is indexed by SCOPUS, WOS is much easier, if not then, we will have to develop robust algorithm equivalent of Google scholar to fetch this data, which is out of scope for this research.

\item Another reason is many Journals shutdown due to various reasons in a very short span of time. Hence any journal needs minimum incubation time up to 3 years to prove its worth. 

\end{enumerate}

     \section{Algorithmic Overview:} This section discusses the basic steps taken to compute internationality and to generate granular clusters. \\
\textbf{Step 1:} Collect data (algorithms to extract data are shown in section 5.3)\\ 
\textbf{Step 2:} Pose internationality as a score: "y" as response variable.\\
\textbf{Step 3:} Model y = $f(x_1,x_2, x_3,.....x_i)$; i= 1,2,.....n
where $x_1, x_2.... x_i$ are the input variables as will be discussed in secton 6.\\
\textbf{Step 4:} (i)Perform down-selection, in case there are too many input variables, some of which could be highly correlated. Otherwise, go back to Step 3.\\
(ii) For simulation and visualization aid use a 3-D down-selection model.
     \begin{center}
      $\displaystyle y = A\prod _{i=1}^2 x_i {^{{\alpha}_i}} $
     \end{center} , obtain "best" estimate of ${{\alpha}_i}$ ; use the best fit values.\\
\textbf{Step 5}: Compute "y" for each category. \\
\textbf{Step 6}: Observe the density and histogram plot. \\
\textbf{Step 7}: Decide on the granularity of internationality into several classes.\\
\textbf{Step 8}: Predict/visualize "variations" in "y" based on small perturbations in $x_1 and x_2 $. \\

The next section details the procedures and algorithms critical for data acquisition from the public domain. These include extracting data for the input parameters required for the model. Additional information such as journal name, country name etc is scraped for building a public repository.

\section{Data Acquisition from Google Scholar}
\label{sec:7}
\subsection{Collection}  There are many advantages of using Google Scholar as a data source because it is free to access, easy to use and quick and comprehensive in its coverage.
     Various studies \cite{5} have also shown that Google Scholar is a serious alternative data source for various reasons. 
     \begin{enumerate}
         \item Not everything published on the internet is counted in Google Scholar:\\
         Google Scholar indexes only scholarly publications. As their website indicates "we work with publishers of scholarly information to index peer-reviewed papers, theses, preprints, abstracts and technical reports from all disciplines of research". Some not scholarly citations, such as student handbooks, library guides or editorial notes slip through.
         \par There might be some overestimation of the number of non-scholarly citations in Google Scholar, for many disciplines this is preferable to very significant and systematic under-estimation of scholarly citations in ISI or SCOPUS. 
         \item Non-ISI publications can be high-quality publications:\\
         There is a misconception that ISI listing is a stamp of quality and one should ignore non-ISI listed publication and citations. 
         \par However, there are a few problems with this assumption. a) ISI has a bias towards Science and English language, b) ISI ignores the majority of publications in the social sciences and humanities as well as engineering and computer science fields.
         
         \item Google Scholar flaws don’t impact citation analysis much:\\
         There is no doubt that the Google Scholar’s automatic parsing occasionally provides us with nonsensical results. However, these errors do not appear to be frequent or important. They do not generally impact the results of author or journal queries much, if at all. 
         \par What is more important is that these errors are random than systematic. In contrast, the commercial databases such as ISI and Scopus have systematic errors that do not include many journals, nor have good coverage of conference proceedings, books or book chapters. Therefore, although it is always a good idea to use multiple data sources, rejecting Google Scholar out of hand because of presumed parsing errors is not rational.
         
     \end{enumerate} 
     In spite of the fact that Google Scholar is an incomprehensible storehouse with uninhibited access, it does not provide an API. Moreover, Google obstructs any computerized web crawling. Subsequently we turned to web scraping. With  occasional time delays included in the script, we could gather the required data from Google Scholar.
     
     \subsection{Organization}  Web scraping is the procedure of consequently gathering data from the World Wide Web. 
          Under this, we plan to develop completely robotized frameworks that can change over whole web website into organized data for further handling. 
          Fig. 4 demonstrates the essential segments of our methodology of web scraping Google Scholar. DOM Parsing is the philosophy which assists the system with retrieving element content created by client-side scripts utilizing undeniable web program controls, for example, the Internet Explorer browser or the Mozilla browser control. 
          These program controls likewise parse web pages into a DOM tree, in light of which program can recover parts of the pages.
     
     \begin{figure}
         \centering
         \includegraphics[width=11cm]{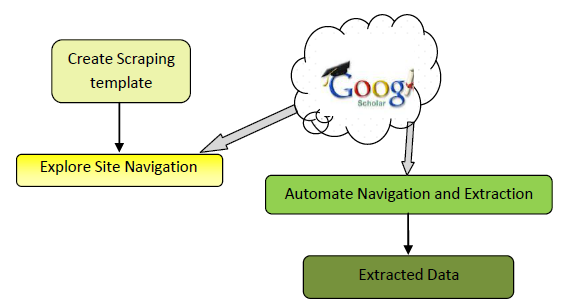}
         \caption{Block diagram of Web Scraping Methodology}
         \end{figure}
     
     \begin{enumerate}
         \item Create Scraping Template: Inspect Element is a developer tool that allows to view the HTML, CSS, and JavaScript that is currently on a web page.    
         On nearly any web page, one can right click and select “inspect element”. This will pull up the developers console to view will the HTML and CSS of the web page.
         Using this tool we explored the Document Object Model (DOM) tree of the Google Scholar Engineering and Computer Science Section.
         Fig. 4 depicts the screen capture of the Inspect Element tool in use.\\
        Left lower panel in the Fig. 4 is the Elements panel used to inspect all elements in the Google Scholar web page (top pane) in one DOM tree. Using this Elements panel one can select any element and inspect the styles applied to it.
        The right lower panel is Styles pane. It shows the CSS rules that apply to the selected element, from highest priority to lowest.
        Styles pane is used to view and change the CSS rules applied to any selected element in Elements panel.
        Following are the various other information access provided in Inspect Element tool.
        \begin{itemize}
        \item Elements: Shows the HTML for the current page
        \item Network: Shows all the GET and POST requests that are made while the developers console is open. One can also identify the requests that are taking the longest to process.
             \item   Sources: Allows to view the JavaScript files (and other files) associated with the page. This is most used for debugging as a web page is being developed, but can be helpful for coding your own JavaScript in Qualtrics as well.
               \item Timeline: The timeline shows where time is invested when a web page is loaded/refreshed. It logs GETs, PUTs, calculations, parsing JavaScript, etc.
               \item Profiles: Also helps see where time is being spent on a page. One can record time spent by function, by JavaScript Object, and by script
               \item Resources: Allows to inspect the resources that are loaded onto a page. (i.e. cookies)
               \item Audits: Analyzes a page as it is loading and then gives suggestions to decrease the load time
               \item Console: This is a JavaScript console where one can try out code as if he/she were coding it for the web page. One can use it to log information about debugging, to test out code snippets, etc.
                
        \end{itemize}

         \begin{figure}
             \centering
             \includegraphics[width=11cm]{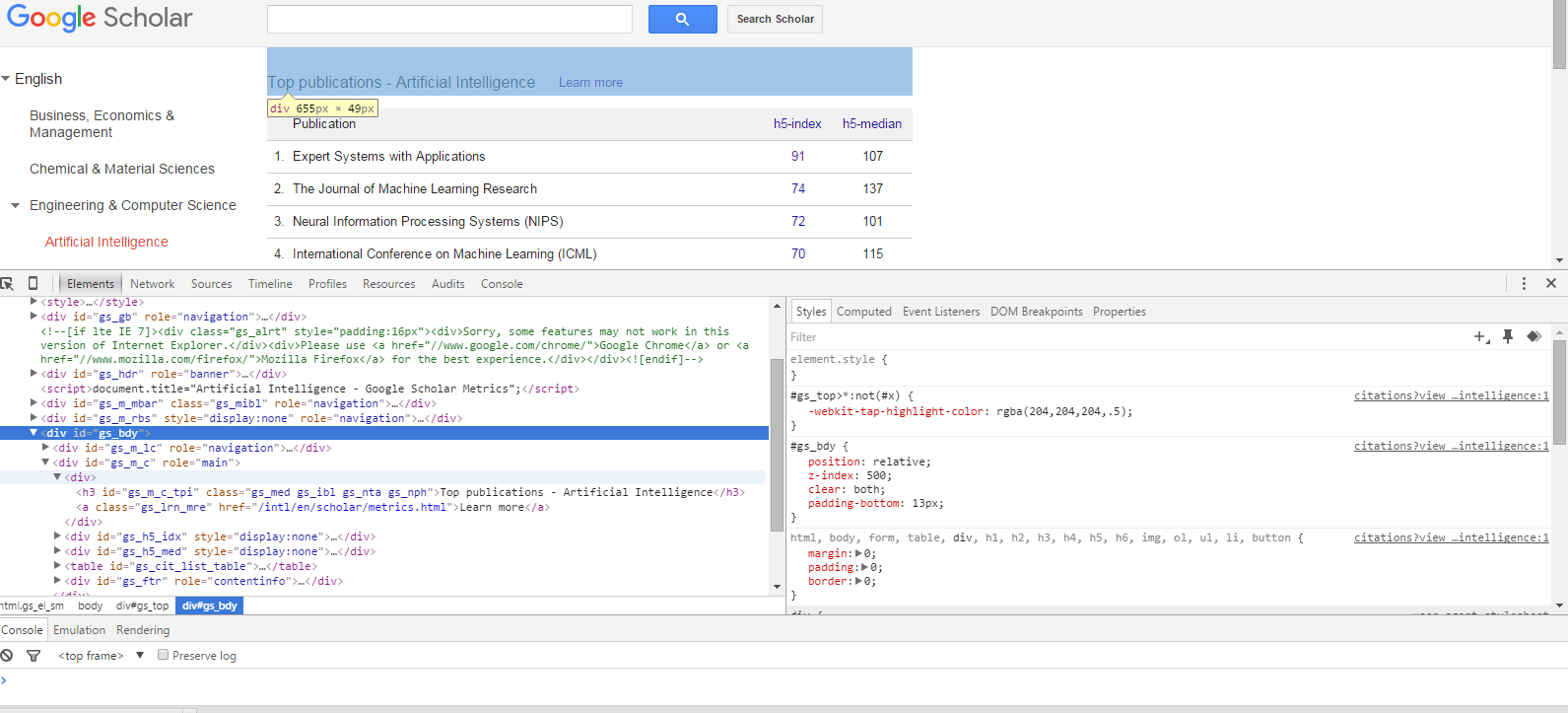}
             \caption{Inspect Element tool usage to explore Document Object Model(DOM).}
               \end{figure}
         
         \item Explore Site Navigation: To further explore and understand the Google scholar site navigation for dynamic URL formulation, we used Beautiful Soul Parser.
        BeautifulSoup parser is also called Elixir and Tonic “The Screen-Scraper’s Friend” \cite{18}
        It uses a pluggable XML or HTML parser to parse a (possibly invalid) document into a tree representation. Beautiful Soup provides methods and pythonic idioms that make it easy to navigate, search, and modify the parse tree. \\
            
         \item Automate Navigation and Extraction: Python is a scripting language which is easy to learn, powerful programming language. We have used the python interpreter library which is freely available in source and binary form for all major platforms.\\
      \end{enumerate}   
         \subsection {Extraction} 
        
        \subsubsection{Algorithm for Feature Extraction - Algorithm 1} Algorithm 1 is for features extraction and internationality index computation for all the listed journals from Google Scholar (data source) under engineering and computer science field.
         Features such as total citations, other-citation, International Collaboration Ratio and SNIP are computed for each one of these journals later on. Also, additional data was obtained from Aminer Citation Network data set\cite{26} based on a paper by Jie Tang et al. \cite{25}. Using web scraping we first extract all the journal names from the source: line 1. Then extract Total Citations count and all the Articles published in each one of these journals: line 3 and 4. Further on we compute the cumulative/averaged parameter values for that journal from the various values extracted for each article: line 5 to 8. The various function calls in these lines are explained ahead in the report under respective algorithms.  the average value for the International Collaboration is computed: line 11. Finally, line 12 and 13 invoke the functions to compute the SNIP and  Internationality Index.\\
        
\begin{algorithm}
              \begin{algorithmic} [1]
                  \State \textbf{Input:} URL link of Google Scholar
                  \State \textbf{Output:}  Features such as International Collaboration Ratio, SNIP, Other-Citations and Internationality Index
                  \State $JNames[]$ = $Fetch\_Journal\_Names\_from\_Google$(Engineering and Computer Science)
                  \For {every journal: $JNames[i]$}
                  \State TotalCites = Get the totalcites value
                  \State Get all the published articles/papers: $X[ ]$
                  \For {every article: $X[i]$}
                  \State $JNames[i]$.Selfcites += compute\_SelfCitations($X[i]$)
                  \EndFor
\State $x_1 = 1- JNames[i].Selfcites/TotalCites$; \textbf{compute other citation quotient} \State $x_2 = compute\_Intl\_Collaboration\_Ratio(JNames[i])/100$; \textbf{compute International Collaboration Ratio}
\State $x_3 = compute\_SNIP(JNames[i])/MaxSNIP$; \textbf{compute SNIP}
\State $x_4 = compute\_NonLocalIQ(JNames[i])$; \textbf{compute NLIQ}
\State $Internationlity\_index = CobbDouglasModel(JNames[i], x_1,x_2,x_3,x_4)$; \textbf{compute JIMI}
\Comment refer section 6 for Cobb-Douglas Model
                  \EndFor
              \end{algorithmic}
              \caption{Driver Algo: Algorithm to extract various features and to compute Internationality Index of Journals}
              \label{algo0}
          \end{algorithm}

         \subsubsection{Journal Name Extraction - Algorithm 2} 
         This algorithm is to extract the journal names for Algorithm 1 to work upon. For the given source's URL, we perform web scraping to first extract all the subcategories of the Engineering and Computer Science field: line 1 and then in turn scrape the 20 journal names listed in each of the web links for these subcategories by dynamically generating the URL addresses using subcategory names: line 2
         Then on we accumulate these scraped journal names in the spreadsheet: line 3
         We successfully extracted about 1160 journal names from all the subcategories listed under Engineering and Computer Science category in Google Scholar. Fig. 5 shows the sample capture of the journal names extracted.
     \begin{algorithm}
                 \begin{algorithmic} [1]
                     \State \textbf{Input:} A html file of Google Scholar web page: $HLINK$
                     \State \textbf{Output:}  List of Journal Names
                     \For {every sub category link in $HLINK: SUBLINK$}
                     \For {every hyperlink in the $SUBLINK: JLINK$} 
                       \State Print $JLINK.gs\_title$ from $ <td>$ tag to spreadsheets
                     \EndFor
                     \EndFor
                     
                 \end{algorithmic}
                     \caption{$Fetch\_Journal\_Names\_from\_Google()$: Algorithm to Extract Journal Names from Google Scholar}
                     \label{algo1}
                 \end{algorithm}

         \begin{figure}
             \centering
             \includegraphics[width=11cm]{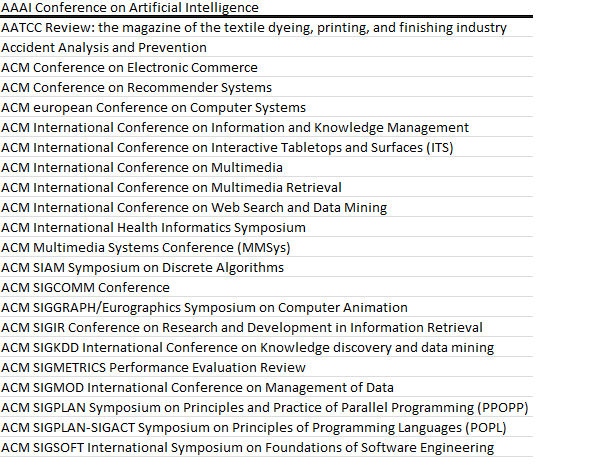}
             \caption{Sample list of Journals extracted into spreadsheet.}
         \end{figure}
         \subsubsection{\textbf{Other-Citation Quotient} Computation - Algorithm 3 } 
         Self-citation is defined as a citation where the citing and the cited paper share at least one author. Other-Citation is the complement of self-citation/total citations, i.e $ 1 - $ self\_citation/total citations. Algorithm 3 provides the skeleton of self-citation computation for an article in a journal. The denominator, total citations, is already computed by parsing web sources. The key to computing Other-Citations Quotient is to calculate self-citations. For this, we first scrape all the cited papers for the input article name (line 1). Then for each one of these cited papers check if it shares at least one common author name with the input article. If true then we increment the self-citation count (lines 3 and 4).  Google Scholar lists a maximum of 1000 cited papers for any listed article. By adding all the individual self-citation counts for every article in a journal, we will get the total self-citations count for a journal (line 5 in Algorithm 1).
         Fig. 6 shows the output from this algorithm which is a raw data extracted to spreadsheet.
         Fig. 7 shows the processed raw data which provided the total self-citations for every listed journal.
                  \begin{algorithm}[H]
                    \begin{algorithmic} [1]
                        \State \textbf{Input:} article/paper name ($P$) from Google Scholar
                        \State \textbf{Output:} self-citation count for article / paper ($P$) 
                        \State Get all citedPapers for article/paper($P$): $citedBy[]$ 
                        \For {Every cited paper: $citedBy[i]$}
                        \If {$P.author\_name$ \textbf{IN} $citedBy[i].author\_names$}
                        \State $IncrByOne(P.SelfCitationCount)$
                        \EndIf
                        \EndFor
                        \State \Return {$SelfCitationCount$}
                    \end{algorithmic}
                    \caption{$compute\_Self\_Citations()$: Algorithm to compute Self-Citation Count}
                    \label{algo2}
                \end{algorithm}

         \begin{figure}[H]
             \centering
             \includegraphics[width=11cm]{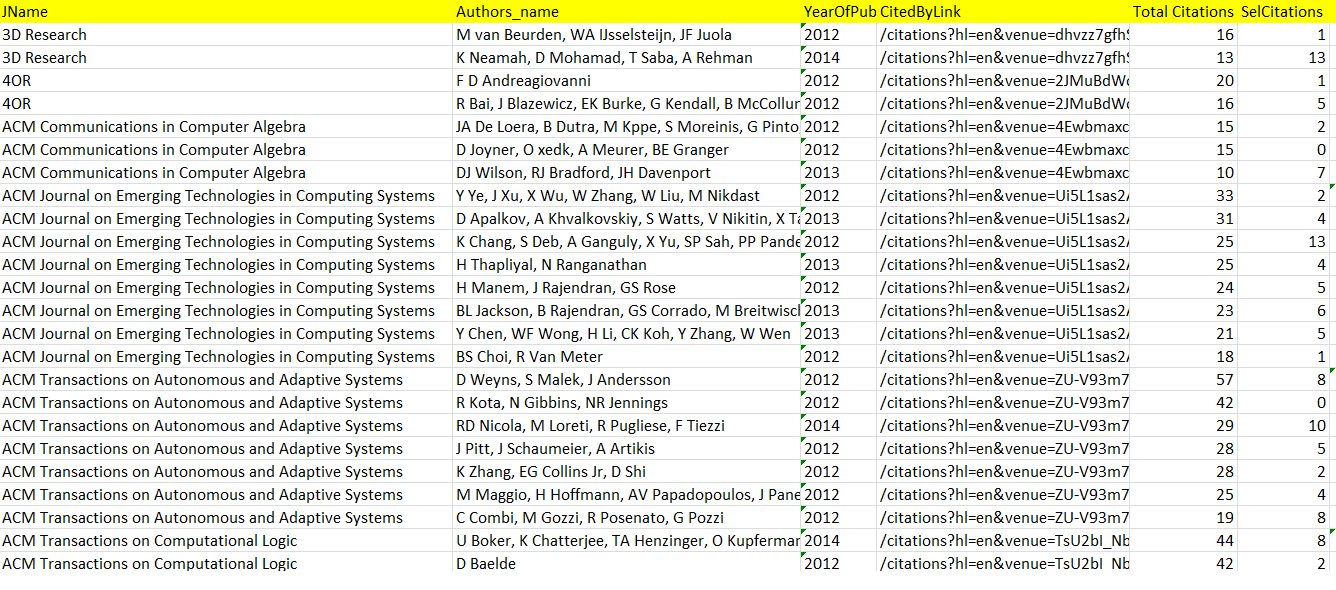}
             \caption{Sample data extracted into spreadsheet.}
         \end{figure}

\begin{figure}
 \centering
            \includegraphics[width=6.5cm]{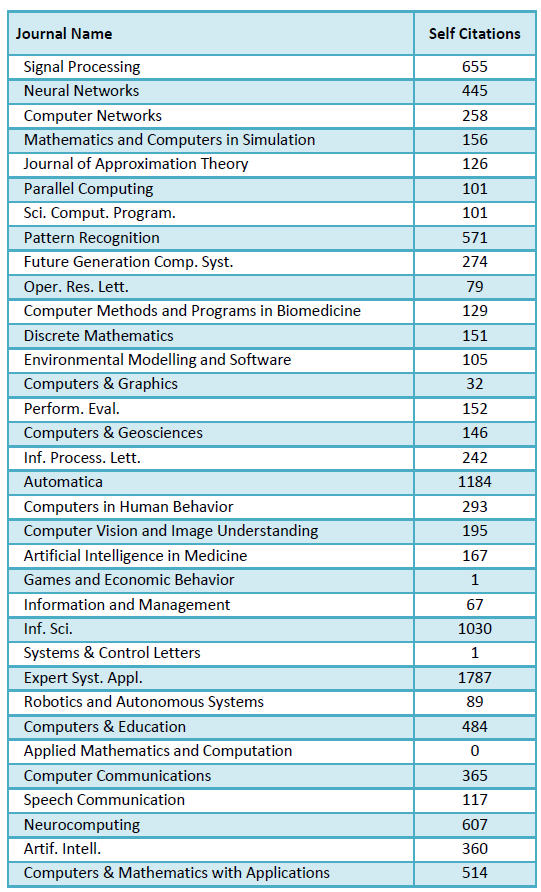}
            \caption{Journals with corresponding self-citation counts.}
            \end{figure}

     \subsubsection{Non-Local Influence Quotient (\textbf{ NLIQ} ) - Algorithm 4 }     
  \par Influence is termed as a factor which causes a paper to be cited by other papers. Non-local refers to the fact that some citations originate from different journals; that is, not from the same journal in which the cited paper is published in. Thus, Non-Local Influence Quotient (NLIQ) is defined as follows,
          \par Let \textbf{A} be the number of citations made from articles in one journal X to articles belonging to a number of different journals.
         Let \textbf{B} be the number of citations from articles in journal X to articles in the same journal, X. Then, for a given journal, we have:
        \par 
        \begin{equation*}
         \textbf{Non-Local Influence Quotient} =  \dfrac{A}{A+B} 
        \end{equation*}
        
\par It must be stressed that \textbf{other-citations} are uniquely different from \textbf{non-local influence}. Namely, an other-citation occurs when a paper cites another paper where no authors are in common. On the other hand, non-local influence is the number of citations made from one paper in a given journal, to a number of different journals - divided by the total number of citations. Section 7 outlines the implications of collaboration and NLIQ towards "internationality" and shows that SNIP values of journals are weakly correlated to their Non-Local Influence Quotient (NLIQ).        

\par Weak non local influence is different from Zero non local influence by definition. So, when a journal has a low NLIQ, journal’s high SNIP will automatically compensate for bringing its Internationality score high enough to be ranked as a good international Journal. 
For example:  Journal of discrete mathematics has low NLIQ, but this low value will be compensated by its high SNIP value when we compute internationality index for it. So, it will still have a good score.
\par We score any Journal with zero non local influence with zero internationality index since it is Journal with no non local influence what so ever.  Within a small community of niche journals; NLIQ =0 implies there is no diffusion. That can’t be good since that implies the articles published in that journal are not able to influence/generate idea strong enough to permeate even within the community, obliviate outside it.
For example :  Astroinformatics, an emerging area; has quite a few journals; JAC, APJ, MNRAS, Astronomy and Space science to name a few. So the concern that only one journal may exist in a specific domain rendering NLIQ of that Journal = 0 is a misplaced appropriation.
\par So when the NLIQ is low (and not zero), the convex combination of JIS and JIMI balances out. Even if NLIQ is low, it doesn’t necessarily make JIMI low, because the presence of other factors will weigh in.  Internationality is not geospatial measure of Journal. We define it as a measure which is devoid of local influence OR non local influence which compliments internationality.
In order to compensate for a low NLIQ in a valid case, the number of journals and articles published in the subject area need to be computed so that, normalization and total number of citation information is also available. This will help in rendering appropriate weights on NLIQ across different subject area.

\algnewcommand\algorithmicforeach{\textbf{for each}}
\algdef{S}[FOR]{ForEach}[1]{\algorithmicforeach\ #1\ \algorithmicdo}

\begin{algorithm}[H]
\caption{$compute\_NonLocalIQ()$: Algorithm to calculate Non-Local Influence Quotient}
\begin{algorithmic}[1]
\State \textbf{Input: }$journal\_name, citation\_database$
\State \textbf{Output: }$NLIQ \enspace of journal\_name$
\State $A\gets 0$ \Comment{external citation count}
\State $B \gets 0$ \Comment{internal citation count}
\State $J\_articles \gets \left[ \enspace \right]$ \Comment{used to store articles in a journal}
\State $count \gets 0$

\ForEach{$article \in citation\_database$} \Comment{get all articles in a journal}
	\If{$article\left[journal\right] = journal\_name$}
		\State $J\_articles[count++] \gets article$
     \EndIf
\EndFor

\ForEach{$article \in  J\_articles$} \Comment{get count of internal, external cites}
	\ForEach{$reference \in article\left[references\right]$}
    	\If{$reference \in ARTICLE\_TYPE$} \Comment{reference is an article}
    		\If{$reference\left[journal\right] \enspace \textbf{!=} \enspace journal\_name$}
        		\State $A \gets A + 1$
        	\Else
        		\State $B \gets B + 1$
        	\EndIf
       \EndIf
    \EndFor
\EndFor
\State $NLIQ \gets A \enspace / \enspace \left( A + B \right)$
\State \Return $NLIQ$
	
\end{algorithmic}
\end{algorithm}

          \subsubsection{International Collaboration Ratio - Algorithms 5, 6, 7} International collaboration accounts for the articles that have been produced by researchers from several countries. 
In order to compute this parameter, we first extracted the country information of the journal and then the author affiliations for each one of the published articles in that journal. Every author's country is matched with the country of publishing journal. Ratio is calculated on the basis of weights assigned to different combination of authors affiliation and origin of the publishing journal.
Algorithm 5 is for collecting the country information of the journals.  Fig. 8 shows the sample of country names of a few listed journals. 
\par Algorithm 6 shows computation of International Collaboration Ratio of a journal. In international collaboration we look for collaboration between two or more scholars with affiliating institutes in different countries. A person with multiple affiliations then, we will pick primary/first listed affiliation during international collaboration computation. 
\par We scrape the data of multiple affiliations (Algorithm 7) from the websites but consider only the primary (first listed) institute in the computation of international collaboration.  Primary reason for this approach is the occurrences of nexus of dummy affiliations in middle-east countries \cite{35}.
 As Bhattacharjee (2011) \cite{38} reported some years ago in Science, Saudi Arabian universities offer highly cited researchers contracts in which the researchers commit themselves to listing the Saudi Arabian University as a further institution in publications (or on highlycited.com). In return, the researchers receive an adjunct professorship which is connected with an attractive salary and a presence at the University of only one or two weeks per year (for teaching duties on site).  Gingras (2014a) \cite{36, 37} names the added institutions as  “dummy affiliations”, with no real impact on teaching and research in universities, allow marginal institutions to boost their position in the rankings of universities without having to develop any real scientific activities.” 
\par Secondly, the best way could have been to have the author provide weights (or number of weeks) for each affiliation. Nonetheless this approach is out of scope in our research work. Therefore, we will consider the primary/first listed affiliation only for further calculation.
\par NOTE: Typically authors are affiliated to more than one institute in same country. A few examples include Harvard University and Center for Astrophysics. NewYork University and Courant Institute of Mathematical Sciences, University of Texas Arlington and Automation and Robotics research institute, Cambridge University and the Institute of Astrophysics. In the western world, land grant institutions under a bigger university setup is very common triggering multiple affiliations to a bunch of faculty associated with the larger university setup. In such cases multiple affiliations do not imply different institutions.

                \begin{algorithm}
                    \begin{algorithmic} [1]
                        \State \textbf{Input: }List of Journal names from Algorithm 1: $J$
                        \State \textbf{Output:} Country information of the Journals 
                        \For {Every journal\_name: $JNames[i]$}
                        \State $Fetch URL http://www.scimagojr.com/journalsearch.php?q="+journal\_name+"\&tip=jou" $
                        \State $write\_to\_spreadsheet(forall('div', {'id': 'derecha\_contenido'}) $
                        \EndFor
                    \end{algorithmic}
                    \caption{$Country\_info(Journal\_name)$: Algorithm to fetch country information}
                    \label{algo2}
                \end{algorithm}

            \begin{figure}[H]
             \centering
             \includegraphics[width=11cm, height=9cm]{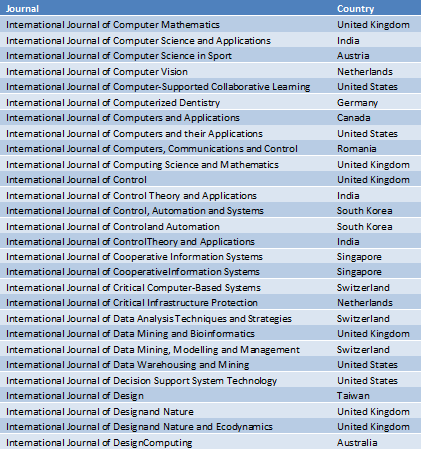}
                  \caption{Sample list of the Journals with country name }
                \end{figure}

\begin{algorithm}[H]
                    \begin{algorithmic} [1]
                        \State Input: Journal Name: $J$ 
                        \State URL to all the articles in that Journal : $J.all\_articles\_url[]$
				        \State Country information of the Journal: $J.contryName$
                        \State Output: \%international collaboration ratio of Journal: $J$
                       \Comment{ 	Compute the internationality weight of an article Based on the combination (Eg: out of 5 authors 2 are from same rest from other) deduce the weight of the article from a predefined values for a given combination,
Eg: For all authors from different countries weight=1, 
    For all authors from same country weight = 0,
    For n/2 authors from one country and n/2 from others weight=0.5 }       
    \State $authAffs = []$
    \For {Every article in $J.all\_articles\_url[i]$ }
    \State $Authors\_Affiliation \leftarrow Fetch\_Author\_Affiliations(article)$ \Comment Algorithm 7
        \State $authAffs.append(read\_author\_name\_and\_first\_affiliation information(Author\_Affiliation))$
       \Comment {Generate 2D array of i:author name, j:country name}

    \State $iNtrNationality\_wt[i]= compute\_wt(article) $
    \EndFor
 
\State $J.iNtrNational[][]$
\Comment {Create one big matrix for a journal where i:country names, j:author names}

\For {every i in $authAffs$}

\If {$Country\_of(i['Affiliation']) == J.countryName) $ } 
\Comment {if author's country same as   Journal's country then make entry = 0}
\State   ${J.iNtrNational[Country\_of(i['Affiliation'])][i['Author']] = 0}$
\Else 
\State  $J.iNtrNational[Country\_of(i['Affiliation'])][i['Author']] = 1$ 
\EndIf
\EndFor
\State x = Ratio of(Number of 0's and Number of 1's in $J.iNtrNational[][]$)
\State y = cumulative weights($iNtrNationality\_wt[i]$)
\State \Return $(\%international\_collaboration = \alpha x + (1-\alpha) y)$ \Comment $\alpha$ is a weight deduced from cross correlation
\end{algorithmic}
                    \caption{$Intl\_Collaboration\_Ratio(JNames[i])$: Algorithm to compute international collaboration ratio of a Journal}
                    \label{algo2}
                \end{algorithm}

Algorithm 7 illustrates steps to fetch author affiliations of an article. An article URL is scraped to obtain Author name and respective affiliations. We extract all the affiliations in case of multiple affiliations for an author. results of one such article are in the as shown below

\begin{algorithm}[h]
 \begin{algorithmic} [1]
 \State \textbf{Input:} Link to the article from algorithm 5: $article\_URL$
 \State \textbf{Output:} Author names and respective Affiliations
 \State $authors[] \longleftarrow  scraped\_author\_names(article\_URL)$ 
 \State $list = []$  \Comment list of dictionaries
 \For { every $author$ in $authors[]$ }
 \State $dictionary\_element = \{'Author': author\} $
 \State $count  = 1$ 
 \For {every $affiliation$ of $author$ }
 \State $dictionary\_element.update\{'count':$affiliation$\} $
 \State $count = count + 1 $ \Comment First, Second, Third Affiliations 
 \EndFor
\State $list.append(dictionary\_element)$
 \EndFor
 \State \Return list
 \end{algorithmic}
 \caption{$Fetch\_Author\_Affiliations(article)$: Algorithm to fetch author affiliations information for the article}
\label{algo2}
\end{algorithm}

\textbf{Data cleansing:} The scraped raw data is not ready for use until it is cleansed further and pre processed. We did following operations to extract the useful data while scraping and post scraping. 
\begin{itemize}
\item Remove extra spaces such as tab, newline etc.
\item Strip/truncate unwanted characters suffixed and postfixed such as \lq \# \rq for the reference ID of the Institution in the html tags.
\item Encode text using UTF-8 encoder in order to take care of the Unicode characters in the extracted raw data while displaying.
\item The Institution Name is a complete address so we   pre-process this address string to extract just the country name (in Algorithm 6) for International Collaboration ratio computation, which is solely based on just the country information of the Affiliated Institution.
\end{itemize}

\fbox{
\begin{minipage}{40em}
\{'1': u'Institute of Space Sciences (CSIC-IEEC), UAB, Barcelona 08193, Spain',
 \\'2': u'Institute for Sciences of the Cosmos (ICC), University of Barcelona, Barcelona 08028, Spain',
 \\'3': u'Department of Astrophysical Sciences, Princeton University, Princeton, NJ 08544, USA',
 'author': u'Beth A. Reid'\}
 \\ \{'1': u'Institute of Cosmology and Gravitation, University of Portsmouth, Portsmouth P01 2EG',
 'author': u'Will J. Percival'\}
\\ \{'1': u'Steward Observatory, University of Arizona, 933 N. Cherry Ave., Tucson, AZ 85121, USA',
\\ 'author': u'Daniel J. Eisenstein'\}
\\ \{'1': u'Institute of Space Sciences (CSIC-IEEC), UAB, Barcelona 08193, Spain',
 \\ '2': u'Institute for Sciences of the Cosmos (ICC), University of Barcelona, Barcelona 08028, Spain',
 \\ '3': u'ICREA (Institucio Catalana de Recerca i Estudis Avancats), Passeig Llus Companys 23, 08010 Barcelona, Spain',
 \\ 'author': u'Licia Verde' \}
\\ \{'1': u'Department of Astrophysical Sciences, Princeton University, Princeton, NJ 08544, USA',
 \\ '2': u'Princeton Center for Theoretical Science, Princeton University, Jadwin Hall, Princeton, NJ 08542, USA',
 \\ 'author': u'David N. Spergel' \}
\\ \{'1': u'Max-Planck-Institute for Astronomy, nigstuhl 17, D-69117 Heidelberg, Germany',
 \\ 'author': u'Ramin A. Skibba' \}
\\ \{'1': u'Department of Astrophysical Sciences, Princeton University, Princeton, NJ 08544, USA',
 \\ 'author': u'Neta A. Bahcall' \}
\\ \{'1': u'Department of Physics and Astronomy, The Johns Hopkins University, 3701 San Martin Drive, Baltimore, MD 21218, USA',
 \\'author': u'Tamas Budavari' \}
\\ \{'1': u'Particle Astrophysics Center, Fermilab, PO Box 500, Batavia, IL 60510, USA',
 \\ '2': u'Kavli Institute for Cosmological Physics, Department of Astronomy \& Astrophysics, University of Chicago, Chicago, IL 60637,USA',
\\ 'author': u'Joshua A. Frieman' \}
\\ \{'1': u'Institute for Cosmic Ray Research, University of Tokyo, Kashiwa 277-8582, Japan',
 \\ 'author': u'Masataka Fukugita' \}
\\ \{'1': u'Department of Astrophysical Sciences, Princeton University, Princeton, NJ 08544, USA',
 \\ 'author': u'J. Richard Gott' \}
\\ \{'1': u'Department of Astrophysical Sciences, Princeton University, Princeton, NJ 08544, USA',
 \\'author': u'James E. Gunn'\}
\\ \{'1': u'Department of Astronomy, University of Washington, Box 351580, Seattle, WA 98195, USA',
 \\'author': u'deljko Ivezi' \}
\\ \{'1': u'Department of Astrophysical Sciences, Princeton University, Princeton, NJ 08544, USA',
 \\ 'author': u'Gillian R. Knapp' \}
\\ \{'1': u'Department of Astronomy and Astrophysics, The University of Chicago, 5640 South Ellis Avenue, Chicago, IL 60615, USA',
 \\ 'author': u'Richard G. Kron' \}
\\ \{'1': u'Department of Astrophysical Sciences, Princeton University, Princeton, NJ 08544, USA',
 \\ 'author': u'Robert H. Lupton'\}
\\ \{'1': u'Departments of Physics and Astronomy, University of Michigan, Ann Arbor, MI, 48109, USA',
 \\ 'author': u'Timothy A. McKay' \}
\\ \{'1': u'SUPA; Institute for Astronomy, University of Edinburgh, Royal Observatory, Blackford Hill, Edinburgh EH9 3HJ',
\\ 'author': u'Avery Meiksin' \}
\\ \{'1': u'Institute of Cosmology and Gravitation, University of Portsmouth, Portsmouth P01 2EG',
 \\ 'author': u'Robert C. Nichol' \}
\\ \{'1': u'Los Alamos National Laboratory, PO Box 1663, Los Alamos, NM 87545, USA',
 \\ 'author': u'Adrian C. Pope' \}
\end{minipage}}

\fbox{
\begin{minipage}{40em}
\{'1': u'Lawrence Berkeley National Lab, 1 Cyclotron Road, MS 50R5032, Berkeley, CA 94720, USA',
 \\ 'author': u'David J. Schlegel' \}
\\ \{'1': u'Department of Astronomy and Astrophysics, The Pennsylvania State University, University Park, PA 16802, USA',
 \\ 'author': u'Donald P. Schneider' \}
\\ \{'1': u'Fermilab, PO Box 500, Batavia, IL 60510, USA',
 \\ 'author': u'Chris Stoughton' \}
\\ \{'1': u'Department of Astrophysical Sciences, Princeton University, Princeton, NJ 08544, USA',
 \\ 'author': u'Michael A. Strauss' \}
\\ \{'1': u'Department of Physics and Astronomy, The Johns Hopkins University, 3701 San Martin Drive, Baltimore, MD 21218, USA',
 \\ 'author': u'Alexander S. Szalay' \}
\\ \{'1': u'Department of Physics, Massachusetts Institute of Technology, Cambridge, MA 02139, USA',
 \\ 'author': u'Max Tegmark' \}
\\ \{'1': u'Department of Physics, Drexel University, Philadelphia, PA 19104, USA',
\\ 'author': u'Michael S. Vogeley' \}
\\ \{'1': u'Department of Astronomy, The Ohio State University, 140 West, 18th Avenue, Columbus, OH 43210, USA',
\\ 'author': u'David H. Weinberg' \}
\\ \{'1': u'Department of Astronomy and Astrophysics, The University of Chicago, 5640 South Ellis Avenue, Chicago, IL 60615, USA',
\\ '2': u'The Enrico Fermi Institute, The University of Chicago, 5640 South Ellis Avenue, Chicago, IL 60615, USA',
\\ 'author': u'Donald G. York' \}
\\ \{'1': u'Department of Astronomy, The Ohio State University, 140 West, 18th Avenue, Columbus, OH 43210, USA',
\\ 'author': u'David H. Weinberg' \}
\end{minipage}}

We consider the only the primary affiliation for computation on international collaboration ratio. \\
\fbox{\begin{minipage}{40em}
Author : Beth A. Reid
\\ Affiliation : Institute of Space Sciences (CSIC-IEEC), UAB, Barcelona 08193, Spain
Author: Will J. Percival
\\ Affiliation: Institute of Cosmology and Gravitation, University of Portsmouth, Portsmouth P01 2EG
Author: Daniel J. Eisenstein
\\ Affiliation: Steward Observatory, University of Arizona, 933 N. Cherry Ave., Tucson, AZ 85121, USA
Author: Licia Verde
\\Affiliation: Institute of Space Sciences (CSIC-IEEC), UAB, Barcelona 08193, Spain
Author: David N. Spergel
\\Affiliation: Department of Astrophysical Sciences, Princeton University, Princeton, NJ 08544, USA
Author: Ramin A. Skibba
\\ Affiliation: Max-Planck-Institute for Astronomy, Königstuhl 17, D-69117 Heidelberg, Germany
Author: Neta A. Bahcall
\\ Affiliation: Department of Astrophysical Sciences, Princeton University, Princeton, NJ 08544, USA

Author: Tamas Budavari
\\ Affiliation: Department of Physics and Astronomy, The Johns Hopkins University, 3701 San Martin Drive, Baltimore, MD 21218, USA

Author: Joshua A. Frieman
\\ Affiliation: Particle Astrophysics Center, Fermilab, PO Box 500, Batavia, IL 60510, USA

Author: Masataka Fukugita
\\ Affiliation: Institute for Cosmic Ray Research, University of Tokyo, Kashiwa 277-8582, Japan

Author: J. Richard Gott
\\ Affiliation: Department of Astrophysical Sciences, Princeton University, Princeton, NJ 08544, USA

Author: James E. Gunn
\\ Affiliation: Department of Astrophysical Sciences, Princeton University, Princeton, NJ 08544, USA

Author: Richard G. Kron
\\Affiliation: Department of Astronomy and Astrophysics, The University of Chicago, 5640 South Ellis Avenue, Chicago, IL 60615, USA

Author: Robert H. Lupton
\\Affiliation: Department of Astrophysical Sciences, Princeton University, Princeton, NJ 08544, USA
Author: Timothy A. McKay
\\Affiliation: Departments of Physics and Astronomy, University of Michigan, Ann Arbor, MI, 48109, USA

Author: Avery Meiksin
\\Affiliation: SUPA; Institute for Astronomy, University of Edinburgh, Royal Observatory, Blackford Hill, Edinburgh EH9 3HJ

Author: Robert C. Nichol
\\Affiliation: Institute of Cosmology and Gravitation, University of Portsmouth, Portsmouth P01 2EG

Author: Adrian C. Pope
\\Affiliation: Los Alamos National Laboratory, PO Box 1663, Los Alamos, NM 87545, USA

Author: David J. Schlegel
\\Affiliation: Lawrence Berkeley National Lab, 1 Cyclotron Road, MS 50R5032, Berkeley, CA 94720, USA

Author: Donald P. Schneider
\\Affiliation: Department of Astronomy and Astrophysics, The Pennsylvania State University, University Park, PA 16802, USA

Author: Chris Stoughton
\\Affiliation: Fermilab, PO Box 500, Batavia, IL 60510, USA

Author: Michael A. Strauss
\\Affiliation: Department of Astrophysical Sciences, Princeton University, Princeton, NJ 08544, USA

Author: Alexander S. Szalay
\\Affiliation: Department of Physics and Astronomy, The Johns Hopkins University, 3701 San Martin Drive, Baltimore, MD 21218, USA

\end{minipage}}

\fbox{\begin{minipage}{40em}

Author: Max Tegmark
\\Affiliation: Department of Physics, Massachusetts Institute of Technology, Cambridge, MA 02139, USA

Author: Michael S. Vogeley
\\Affiliation: Department of Physics, Drexel University, Philadelphia, PA 19104, USA

Author: David H. Weinberg
\\Affiliation: Department of Astronomy, The Ohio State University, 140 West, 18th Avenue, Columbus, OH 43210, USA

Author: Donald G. York
\\Affiliation: Department of Astronomy and Astrophysics, The University of Chicago, 5640 South Ellis Avenue, Chicago, IL 60615, USA

Author: Željko Ivezić
\\ Affiliation: Department of Astronomy, University of Washington, Box 351580, Seattle, WA 98195, USA

Author: Gillian R. Knapp
\\ Affiliation: Department of Astrophysical Sciences, Princeton University, Princeton, NJ 08544, USA

\end{minipage}}

\section{Cobb Douglas Model: Internationality Score function}
In economics, Cobb-Douglas production function \cite{35,37} is widely used to represent relationship of outputs to inputs. This is a technical relation which describes the Laws of Proportion, i.e., the transformation of factor inputs into outputs at any particular time period. This production function is used for the first time, to compute the internationality of a journal where the predictor/independent variables, $ x_i, i= 1,2,...n $ are algorithmically extracted from different sources as explained in the preceding section. 
Internationality, $ y $ is defined as a multivariate function of $ x_i, i= 1,2,...n $. Internationality score varies over time and depends on scholastic parameters, subject to evaluations, constant scrutiny and ever changing patterns. \\
Cobb-Douglas function is given by
     \begin{center}
      $\displaystyle y = A\prod _{i=1}^n x_i {^{{\alpha}_i}} $
     \end{center}
      where y is the internationality score, \\
      $x_i$  are the predictor variables/input parameters 
      and $\alpha_i$ are the elasticity
     coefficients. The function has extremely useful properties such as      convexity/concavity depending upon the elasticity's. The properties yield      global extrema which are intended to be exploited in the computation of    internationality.

\noindent A sample Cobb-Douglas production function for two inputs, $x_1$ and $x_2$  and  internationality of journal as output, \(y\),  is written as -
\[ y=A{x_1}^{\alpha}{x_2}^{\beta}\nonumber \]
where:
\begin{itemize} 
\item $0<\alpha,\beta<1$
\item $y$: Internationality of journal
\item $x_1$: International Collaboration (percentage) 
\item $x_2$: SNIP (Source-Normalized Impact per Paper) 
\end{itemize} 
 As explained in the subsequent sections, the sample model is easily extended to accommodate all relevant input/predictor variables  extracted during the acquisition process [ Please refer Section $5$ ].
\subsection{Functional Form}
Here,  \(x_1\) and \(x_2\) values which are modified values of international collaboration and SNIP respectively, are taken into consideration for different journals and using these optimal values for \(\alpha\) and \(\beta\) is computed. In order to have data lying between 0-1, transformations on the data set is performed, which give the final input values for Cobb-Douglas production function. Finally, the two variables along with the elasticity values of \(\alpha\) and \(\beta\) are placed in Cobb-Douglas equation to compute \textit{y}. Following is the algorithm used:

\noindent Algorithm to find optimum values of \(\alpha\) and \(\beta\):\\
1. Input values of \(x_1,x_2\)\\
2. Vary \(\alpha\) for \(x_1\) such that the corresponding \textit{y} reaches it’s maximum value. Similarly compute \(\beta\) values by varying \(x_2\).\\
\noindent In the figure (Fig. 9) below it can be seen that \textit{y} is maximum for \(\alpha\)=0.1 and \(\beta\)=0.1
\begin{figure}[ht]
  \begin{center}
    \includegraphics[width=7cm]{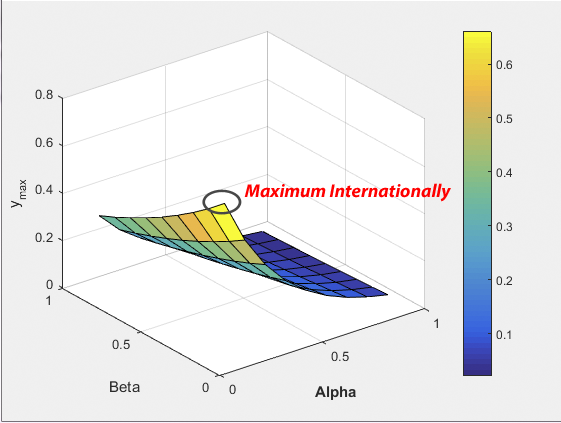}
  \end{center}
  \caption{Optimum values of \(\alpha\) and \(\beta\)}\label{optimum}
\end{figure}

\noindent  Hence, the optimum values of \(\alpha\) and \(\beta\) are 0.1.
Using these values of \(\alpha\) and \(\beta\) in the Cobb-Douglas production function \begin{center}y = $A{x_1}^{\alpha}{x_2}^{\beta}\nonumber$ \end{center} \textit{y} is computed which represents internationality of journal. In the next subsection, regression method is used to compute \(\alpha\) and \(\beta\). The general model is endowed to accommodate any number of factors as proved in section 6.2. However in section 6 and 6.1, two factors are considered as illustration for the 3-D plot. As observed, \(\alpha\) and \(\beta\) are varied to capture the maximum value of "y", internationality of the journal. If the number of input parameters are increased, visualization becomes untenable.

\subsection{Proof of Concept}  The section proves the efficacy of the model for n number of variables/inputs, n being countably finite\\
       As a first exercise, authors have done the simulation for 2 variables and extended to 3. This can be extended to n variables as shown below.\\
   Consider the following production function:
\begin{align*}
&y=\prod_{i=1}^{n}kx_i^{\alpha{_i}}\end{align*}  $n = 4$, $x_1$ to $x_4$ are the input parameters as described below:

\begin{itemize}
        \item $x_1$ : Other-Citations Quotient  =  $1$ - ( self-citations / total citations ) 
        \item $x_2$ : International Collaboration / 100
        \item $x_3$ : SNIP value / maximum SNIP value
        \item $x_4$ : Non-Local Influence Quotient
       
        \end{itemize}
        \par \noindent
        \textbf{Note: }Due to the fact that these input parameters are not just raw numbers but instead defined as quotients having values between 0 and 1 - that is, they are normalized - the Cobb-Douglas model allows for a fair comparison between different subject fields where collaboration and citation trends differ dramatically (such as in Computer Science, the Social Sciences and Mathematics).\\
 
       We have discussed the parameters, $x_1$, $x_2$ in section\textbf{ 5 - Data Acquisition}. This section following the analytical explanation of the Cobb-Douglas model, section $7$ contains discussion on $x_3$ and $x_4$, since both are not merely acquisition oriented but deserve discussion by their own merit.  \\
\newline      
\textbf{\underline{Lemma I}}: Maximum internationality score can be obtained at decreasing returns to scale which is true when -
\begin{align*}
\sum_{i=1}^{n}\alpha_i<1
\end{align*}
where $\alpha_i$ is the $i^{th}$ elasticity of the input variable $x_i$. Consider the following production function:
\begin{align*}
&y=\prod_{i=1}^{n}kx_i^{\alpha{_i}}\end{align*}
To prove:
\begin{align*}
\sum_{i=1}^{n}\alpha_i<1
\end{align*}
Consider the profit function:
\begin{align*}
\pi_n=\prod_{i=1}^{n}kx_i^{\
alpha_i}-\sum_{i=1}^{n}{w_i}{x_i}
\end{align*}
\(w_i\): Unit cost of inputs\\
Profit maximization is achieved when: \(p\frac{\partial f}{\partial x_i}=w_i\). Deriving the condition for optimization:
\begin{align}
pk\frac{\alpha_1}{x_1}&\prod_{i=1}^{n}x_i^{\alpha _i}=w_1\\
pk\frac{\alpha_2}{x_2}&\prod_{i=1}^{n}x_i^{\alpha _i}=w_2\\&.\nonumber \\&. \nonumber \\&. \nonumber \\&. \nonumber\\
pk\frac{\alpha_n}{x_n}&\prod_{i=1}^{n}x_i^{\alpha _i}=w_n
\end{align}
Multiplying these equations with \(x_i\), respectively-
\begin{align}p{\alpha _1}\prod_{i=1}^{n}kx_i^{{\alpha _i}}&=w_1x_1\Rightarrow p\alpha_1y=w_1x_1\\
p{\alpha _2}\prod_{i=1}^{n}kx_i^{{\alpha _i}}&=w_2x_2\Rightarrow p\alpha_2y=w_2x_2\\&. \nonumber \\&. \nonumber \\&. \nonumber \\&. \nonumber\\
p{\alpha _n}\prod_{i=1}^{n}kx_i^{{\alpha _i}}&={w_n}x_n\Rightarrow p{\alpha_n}y={w_n}x_n
\end{align}
Dividing equations (5) .. (6) by (4), following equations are obtained:
\begin{align*}
x_2&=\frac{\alpha_2}{\alpha_1}\frac{w_1}{w_2}x_1\\
x_3&=\frac{\alpha_3}{\alpha_1}\frac{w_1}{w_3}x_1\\ &.\\ &.\\
x_{n-1}&=\frac{\alpha_{n-1}}{\alpha_1}\frac{w_1}{w_{n-1}}x_1\\
x_n&=\frac{\alpha_n}{\alpha_1}\frac{w_1}{w_n}x_1
\end{align*}
Substituting these values of\(x_i\) in equation (1),
\begin{align*}
&pk\frac{\alpha_1}{x_1}\prod_{i=1}^{n}x_i^{\alpha _i}=w_1\\
\Rightarrow &pk\alpha_1x_1^{\alpha_1-1}\left(\frac{\alpha_2}{\alpha_1}\frac{w_1}{w_2}x_1\right)^{\alpha_2}\left(\frac{\alpha_3}{\alpha_1}\frac{w_1}{w_3}x_1\right)^{\alpha_3}....\left(\frac{\alpha_{n-1}}{\alpha_1}\frac{w_1}{w_{n-1}}x_1\right)^{\alpha_{n-1}}\left(\frac{\alpha_{n}}{\alpha_1}\frac{w_1}{w_n}x_1\right)^{\alpha_{n}}=w_1\\
\Rightarrow &pk{x_1}^{\left(\alpha_1+\alpha_2+....+\alpha_n\right)-1}\alpha_1^{1-\left(\alpha_2+\alpha_3+....+\alpha_n\right)}\alpha_2^{\alpha_2}....\alpha_n^{\alpha_n}w_1^{-1+\left(\alpha_2+\alpha_3+....+\alpha_n\right)}w_2^{-\alpha_2}....w_n^{-\alpha_n }=1\\
\Rightarrow&x_1=\left(pk\alpha_1^{1-\left(\alpha_2+\alpha_3+....+\alpha_n\right)}\alpha_2^{\alpha_2}....\alpha_n^{\alpha_n}w_1^{-1+\left(\alpha_2+\alpha_3+....+\alpha_n\right)}w_2^{-\alpha_2}....w_n^{-\alpha_n}\right)^\frac{1}{1-\left(\alpha_1+\alpha_2+....+\alpha_n\right)}
\end{align*}
Performing similar calculations following values of \(x_i, \left(i>=2\right)\) are obtained,
\begin{align*}
x_2&=\left(pk\alpha_2^{1-\left(\alpha_1+\alpha_3+....+\alpha_n\right)}\alpha_1^{\alpha_1}....\alpha_n^{\alpha_n}w_2^{-1+\left(\alpha_1+\alpha_3+....+\alpha_n\right)}w_1^{-\alpha_1}....w_n^{-\alpha_n}\right)^\frac{1}{1-\left(\alpha_1+\alpha_2+....+\alpha_n\right)}\\ &.\\ &.\\
x_n&=\left(pk\alpha_n^{1-\left(\alpha_1+\alpha_2+....+\alpha_{n-1}\right)}\alpha_1^{\alpha_1}....\alpha_{n-1}^{\alpha_{n-1}}w_n^{-1+\left(\alpha_1+\alpha_2+....+\alpha_n\right)}w_2^{-\alpha_2}....w_{n-1}^{-\alpha_{n-1}}\right)^\frac{1}{1-\left(\alpha_1+\alpha_2+....+\alpha_n\right)}
\end{align*}
Substituting values of \(x_i\) in production function,
\begin{align*}
y=\left(kp^{\left(\alpha_1+\alpha_2+....+\alpha_n\right)}\alpha_1^{\alpha_1}\alpha_2^{\alpha_2}....\alpha_n^{\alpha_n}w_1^{-\alpha_1}w_2^{-\alpha_2}....w_n^{-\alpha_n}\right)^\frac{1}{1-\left(\alpha_1+\alpha_2+....+\alpha_n\right)}
\end{align*}
y increases in price of its output and decreases in price of its inputs iff:
\begin{align*}
1-\sum_{i=1}^{n}\alpha_i>0\\
\sum_{i=1}^{n}\alpha_i<1
\end{align*}
Therefore decreasing returns to scale, is validated.

\subsection{Proof of Concavity of Cobb-Douglas function using Hessian Matrix}\
This section proves that the Cobb-Douglas \cite{36,37} production model is concave in nature and hence a maximum internationality score can be found at a particular value of input factors which in this case are international collaboration and SNIP value. The concavity of the function is proved by showing that the Hessian Matrix of the function is negative semi-definite.\\
\noindent{\textbf{Definition:}}
\begin{enumerate}
\item Suppose \(f\in C^2\), U is an open curve set, then \(f:U\subset R^n \rightarrow R\) is concave/strictly concave iff the Hessian Matrix \(D^2f(x)=H\) is negative semi-definite/ negative definite \(\forall x\in U\).\\\(C^2\): Class of continuous and second order differential functions[11].
\item Let S be a convex set [12]; \(x_1,x_2\) be any two points in S; then a function \(f:S \subset R_n \rightarrow R\) is concave if, 
\begin{align} (1-\lambda)f(x_1)+\lambda f(x_2)\leq f((1-\lambda)x_1+\lambda x_2); \hspace{1cm}    \lambda \in [0,1]\nonumber \end{align}
\item Constant and Decreasing Returns to scale: \cite{14}  In the phase of constant returns, an increase in one input may yield an increase in corresponding output in the same proportion. The 3D plots obtained are concave.\\
Whereas, In decreasing returns to scale the deployment of an additional input will result into increase in output but at a diminishing rate or lower ratio.
\end{enumerate}

\noindent{\textbf{Lemma II:} \(f\in C, U\subset R;U\) is a convex, open set,  \(f:R \rightarrow R, f\) is a concave iff } 
\begin{align} f(x+\theta)\leq f(x)+\nabla f(x)\theta; \;\; \;\; \forall\; \;  \theta \in R^{N}; x+\theta \in A; \nonumber \end{align}
C: Class of continuous and first order differential functions, \\ 
\textbf{Proof:} Using the definition of concave functions;
\begin{align*} \hspace{5mm} &f(\alpha (x+\theta)+(1-\alpha)x)\geq \alpha f(x+\theta)+(1-\alpha)f(x) \nonumber \\
&\Rightarrow f(x+\alpha \theta)-f(x)\geq\alpha(f(x+\theta)-f(x)) \nonumber\\ 
&\Rightarrow f(x)+\frac{f(x+\alpha \theta)-f(x)}{\alpha}\geq f(x+\theta)\\ 
&\Rightarrow f(x)+\nabla f(x)\theta \geq f(x+\theta) \;\;\;\; as\;\;\; \alpha \rightarrow 0 \end{align*} 
\\
\textbf{Theorem 1:}\(f\in C^2;x\in R; f:R^2 \rightarrow R\) is concave iff the Hessian Matrix, \(H\equiv D^{2}f(x)\) is negative semi-definite \(\forall x\in U\). [necessary and sufficient condition for concavity]\\ \\
\textbf{Proof:} \(f\) is concave, for some\(\;x\in U\) and some \(\theta \neq 0\), consider the Taylor expansion;
\begin{align}f(x+\alpha \theta)=f(x)+\nabla f(x)(\alpha \theta)+\frac{{(\alpha\theta)}^{2}}{2} D^2f(x+t\theta)\; \; for\; some\; 0<t<\theta \nonumber \end{align}
By lemma;
\begin{align}\frac{{(\alpha\theta)}^2}{2} D^2f(x+t\theta) \leq 0 \nonumber \end{align}
Consider an arbitrary \(\alpha \rightarrow 0,\)\; and\; \(t \rightarrow 0\)
\begin{align}\theta^2 D^2 f(x)\leq 0 \Rightarrow D^2f(x) \leq0 \Rightarrow H\; is\; negative\; semi-definite. \nonumber 
\end{align} 
\subsection{Implications of Theorem 1:}
Cobb-Douglas is concave for conditions on elasticity, thus for such values of elasticity, the Hessian Matrix of the  function is negative semi-definite and therefore concave and attains a global maxima.\\ 
Now consider, the Cobb-Douglas function;$ f(x_1,x_2)=k x_1^{\alpha}x_2^{\beta}$ with $k,\alpha,\beta>0$ for the region $x_1>0$ and $ x_2>0 $
$$H=\begin{bmatrix}\alpha(\alpha-1)kx_1^{\alpha-2}x_2^{\beta}&\alpha \beta kx_1^{\alpha-1}x_2^{\beta-1}\\\alpha \beta kx_1^{\alpha-1}x_2^{\beta-1}&\beta(\beta-1)kx_1^\alpha x_2^{\beta-2}\end{bmatrix}$$
First order principal minors \cite{13} of H are:
$$M_1=\alpha(\alpha-1)k{x_1}^{\alpha-2}{x_2}^{\beta};\; \; \; \; \; \; M_1'=\beta(\beta-1)k{x_1}^{\alpha}{x_2}^{\beta-2}$$
Second order principal minor is:
$$M_2=k\alpha \beta x_1^{2\alpha-2}{x_2}^{2\beta-2}[1-(\alpha+\beta)]$$
H must be negative semi-definite, this implies \(f(x_1,x_2)\) is concave.\\ This will happen if \(M_1\leq0,\; \; M_1'\leq0 \;\;and\; \; M_2\geq0\)\\ For decreasing and constant returns to scale: \(\alpha+\beta\leq1\), therefore 
\begin{align} & \alpha \leq 1, \beta < 1 \nonumber \\
& \Rightarrow (\alpha -1) \leq 0 \nonumber \\
& \Rightarrow M_1 \leq 0 \nonumber \\
& (1-(\alpha+\beta))\geq 0 \nonumber \\ 
& \Rightarrow M_2 \geq 0 \nonumber \end{align}
Both conditions for concave function are satisfied by decreasing and constant returns to scale. Therefore, \(f(x_1,x_2)\) is concave, if
\begin{align} \alpha \geq 0,\beta \geq 0, \alpha+\beta \leq 1\nonumber \end{align}
\textbf{Significance of concavity}:\\ 
The extrema of the function, \(f(x,y)\) used to model "internationality" is useful in finding a global maximal value of the "internationality" indicator. The modeling paradigm is based on the fact that, there exists a maximum internationality score and the score/values in the neighborhood could be classified as the levels of internationality. It is, in this context, we explore if the maxima given by the concave function, i.e.Cobb-Douglas is the global maxima.\\

\textbf{ Theorem 2: Global maxima result:} \\ 
Let \(f(x_1,x_2)=kx_1^{\alpha}x_2^{\beta}: U\subset R^2\rightarrow R\)
be concave function on U; U is an open convex set; the critical point, \(x^{\ast}\) is a global maximum.\\ \\
\textbf{Proof:}	\(x^{\ast}\) is a critical point. Therefore; 
\(Df(x^{\ast})=0\) [D: first order partial derivative]\\
Using a well known result about concave functions;\\
\(f:R^2\rightarrow R\) is concave iff \(f(x_2)-f(x_1)\leq Df(x_1)(x_2-x_1)\;\; \forall x_1,x_2\in U\);\\ Therefore; \\
\(f(x_2)-f(x_1) \leq \frac{\partial f(x_1)}{\partial (x_1)}(x_2'-x_1')+...\frac{\partial f(x_2)}{\partial (x_2)}(x_2^2-x_1^2)\) \\
Since,\\ 
\(Df(x^\ast)\equiv 0\) using the inequality\\
\(f(x_2)-f(x^\ast)\leq Df(x^\ast)(x_2-x^\ast)\) 
        \(\Rightarrow f(x_2)\leq f(x^\ast)\; \; \forall x_2\in U\)
        
        \begin{figure}[htbp]

  \begin{subfigure}{0.5\textwidth}
  \centering
  \includegraphics[width=0.8\linewidth, height=4cm]{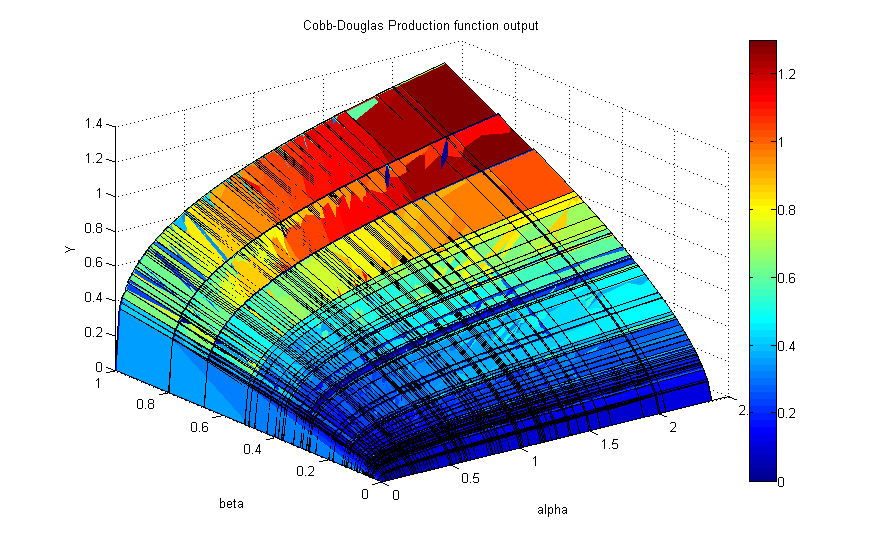} 
  \label{fig:subim1}
  \end{subfigure}
  \begin{subfigure}{0.5\textwidth}
    \centering
  \includegraphics[width=0.8\linewidth, height=4cm]{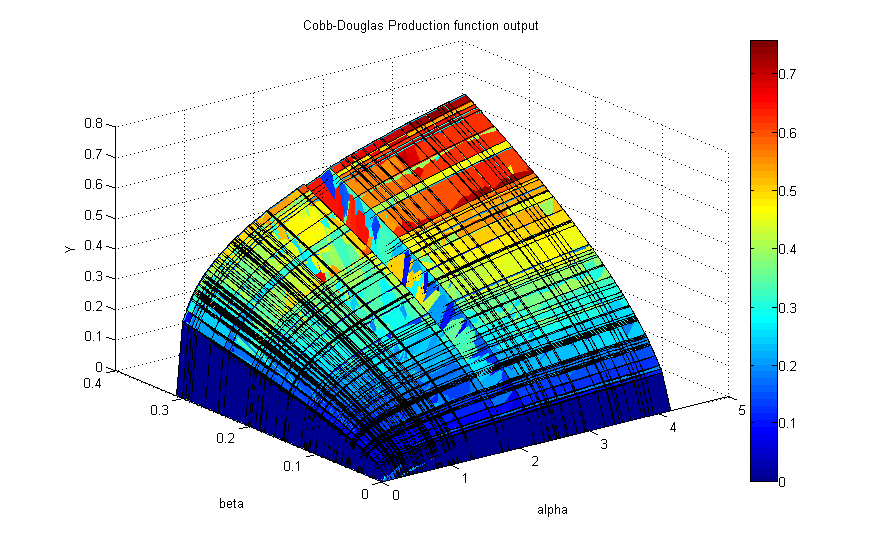}
  \label{fig:subim2}
  \end{subfigure}
  
  \begin{subfigure}{0.5\textwidth}
    \centering
 \includegraphics[width=0.8\linewidth, height=4cm]{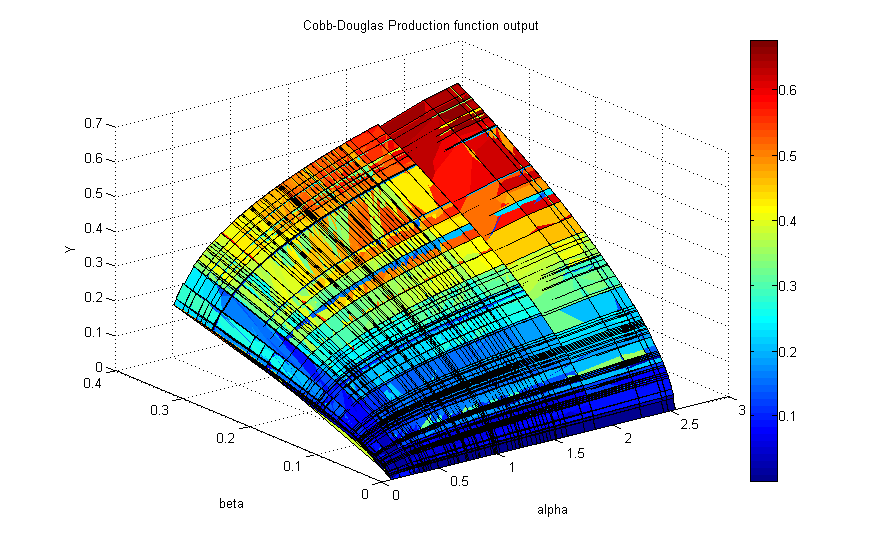} 
    \label{fig:subim1}
    \end{subfigure}
    \begin{subfigure}{0.5\textwidth}
      \centering
    \includegraphics[width=0.8\linewidth, height=4cm]{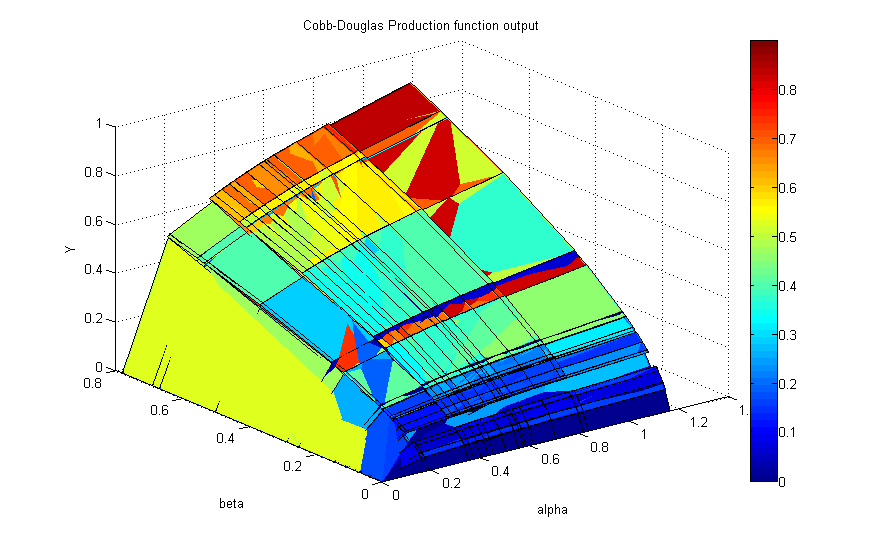}
    \label{fig:subim2}
    \end{subfigure}
    \caption{y, internationality values from Cobb-Douglas Production function at various instances \cite{27} }
  \label{fig:image2}
  \end{figure}
        
\textbf{Note:}
\begin{enumerate}
\item The functional modeling, \(f(x_1,x_2)=kx_1^{\alpha}x_2^{\beta}\) may be extended to \(f(x_1,x_2 \dots x_n)=k\prod_{i=1}^{n} x_i^{\alpha_i}\); in which case \(f:U\subset R^n\rightarrow R\) \& the global maxima holds.
\item U doesn't necessarily be open, the global maxima is guaranteed to be on the closed set as well, since the search for global maxima is allowed on the boundary.
\item Any Cobb-Douglas function is quasi concave. 
\item The values of elasticity are computed by using \textbf{fmincon} command in Matlab. These elasticity values are the exponents in the expression, \(f(x_1,x_2 \dots x_n)=k\prod_{i=1}^{n} x_i^{\alpha_i}\); the function \textbf{fmincon} is a built-in convex optimization tool in MATLAB and corroborates \textbf{Lemma II} proved above.
\end{enumerate}

A 3-D graph of Cobb Douglas function with output measured along the vertical axis is shown in Fig 10. The graph is a part of an AVI file whose frame are created in MATLAB to demonstrates the quasiconcave nature of Cobb Douglas model and to show how y reaches its maximum value at certain input values of $x_1$, $x_2$ $\alpha$ and $\beta$. 
The Matlab code can be viewed on GitHub, Appendix I \cite{32}. The input parameters ($x_1$, $x_2$) are SNIP and other-citations/total-citations and the elasticity coefficients ($\alpha$, $\beta$) are taken along X, Y coordinates. Lower values of y is indicated in blue region which increases and at certain values of elasticity coefficients, reaches to its maximum as marked in red. This is a sample representation and can't include more than two input parameters.\\

  \section{SNIP and Non-Local Influence Quotient, a new metric definition}
          \par The authors have taken a four-pronged approach to thoroughly validate the use of SNIP and NLIQ in the Cobb-Douglas model. First, we shall show the merits of Source-Normalized Impact per Paper (SNIP) over Impact Factor (IF) \cite{28}. Then, the algorithm used to compute SNIP is described and verified with a sample data set. The third sub-section looks at citation patterns with increased granularity; namely inter-journal and intra-journal collaboration which will help show that SNIP is a good indicator of collaboration at the journal level. Lastly, we will look at Non-Local Influence Quotient (NLIQ) described in section 5.3.4 and justify its usage in calculating the internationality of a journal and show why SNIP alone should not be used to determine the relative ranks of journals in academia.
                      
\subsection{Comparison of SNIP and IF}

          \par 
          \par Now, Source-Normalized Impact per Paper (SNIP) measures a source's contextual citation
          impact. It takes into account characteristics of the source's subject field, especially the
          frequency at    which authors cite other papers in their reference lists, the speed at which
          citation impact grows, and the extent to which the database used in the evaluation covers
          the field’s documents. SNIP is the ratio of a source's average citation count per paper, and the
          citation potential of its subject field. It aims to allow direct comparison of sources in different
          subject fields due to the subject-field normalization that takes place in calculating it.
          \par The impact factor (IF) of an academic journal is a measure reflecting the average number
          of citations to recent articles published in that journal. In any given year, the impact factor of a
          journal is the average number of citations received per paper published in that journal during
          the two or five preceding years.

           \par \textbf{SNIP offers several advantages:}
              \begin{enumerate}
              \item \textbf{Openly Available and Greater Coverage}
               How IF is calculated and the source database for citations is known only to Thomson
              Reuters (ISI Web of Science) which means journals not present in their database are not
              assigned an IF value. Also, not all journals indexed by them are provided an IF. This
              disallows researchers from comparing journals which are not indexed.
              \par Scopus, on the other hand, provides journal metrics values to all peer-reviewed 
              journals  
              indexed in their database which is comparably larger. Furthermore, SNIP can be calculated 
              from any Open Access journal using the white paper describing the calculation of SNIP. This 
              allows one to compare journals, however, the types of citations taken into account from 
              Open 
              Access journals must be kept in mind to give as fair a comparison as possible.

              \item \textbf{Subject Field Normalization}
              Life Sciences have a much higher IF as compared to Mathematical journals due to the 
              differences in citation behavior between the two fields.The quality of a journal cannot be 
              derived from its Impact Factor. Due to the fact that SNIP inherently normalizes for 
              differences in citation practices across subject fields, comparison of the ‘prestige’ of two 
              journals belonging to difference subject fields is possible.
     
              \item \textbf{Citation Window}
              SNIP has an ideal citation window, in the authors’ opinion. A three-year citation window 
              allows fields that move at a slower pace to be compared with those that advance fairly 
              rapidly, in as fair a manner as possible. Whereas the 2-year IF and 5-year IF only favor 
              one or the other.
     
              \item \textbf{More Difficult to Game the System}
             A journal’s impact factor is derived from citations of all types of content - including non-peer 
              reviewed material such as editorials. On the other hand, SNIP is derived only from citations 
              of peer-reviewed content and directed to peer-reviewed content, which makes it much more 
              difficult to game the system as the content goes through some form of scrutiny vis-a-vis 
              editorials.
              
              \par Further, given the dramatic increase in predatory journals in recent years who merely 
              charge a fee for publishing an author’s paper albeit with deceitful tactics; and their 
              the proportional increase in their IF values - it is clear that IF is not a suitable metric for 
              measuring the `prestige' of a journal.
\subsection{Algorithm to Compute SNIP}

\par As shown in the original SNIP indicator designed by Henk F. Moed \cite{29} the SNIP indicator is defined as the ratio of a journal's raw impact per paper (RIP) and a journal's database citation potential in its subject field (DCP), that is the RIP value of a journal equals the average number of times the publications of that journal were cited in the three years the year of analysis. For example, if 200 publications were present in a journal from 2009 to 2011 and if these publications were cited 400 times in 2012, the RIP value of the journal for 2012 would be 400 / 200 = 2. In calculating RIP, both citing and cited publications are included only if they have the Scopus document type article, conference paper or review - i.e peer reviewed material. RIP is similar to journal impact factor (IF), although RIP uses three instead of two years of cited publications and only includes citations to the previously mentioned document types. RIP does not account for differences in citation practices among different journals.
\par The DCP value of a journal is equal to the average number of references in the publications belonging to the journal's subject field, where the average is calculated as the arithmetic mean. By finding the ratio of a journal's RIP to it's DCP, we can compare journals belonging to two different fields in a more fair manner. Algorithm 8 shows how to calculate the same.

Although there are certain differences between the original SNIP indicator \cite{29} and the revised SNIP indicator, the authors decided to forgo the latter. This is because an empirical analysis was done between the two and Ludo et al. stated that "from an empirical point of view the differences between the original SNIP indicator and the revised one are relatively small" \cite{12}.

\begin{algorithm}
			\caption{$compute\_SNIP( cites[][], Jpub[], Jsize )$ : Algorithm to calculate SNIP}\label{alg:SNIP}
		\begin{algorithmic}[1]
            \State \textbf{Input:} Database of cites (cites[][]) made to publications of journal J (Jpub[] with Jsize publications) in year X to all documents (article, conference paper or review) in the three years preceding X
            \State \textbf{Output:} SNIP value for journal J in year X

\State $journal \gets \textit{Jname}$
\State $year \gets \textit{read year\_to\_be\_computed\_for}$
\State $citation\_count \gets \textit{0}$
\ForAll{paper in Jpub}
	\State $citation\_count \gets \textit{citation\_count + count of papers published in year - 1, year - 2, year - 3}$
    \State $num\_papers \gets \textit{num\_papers + 1}$
\EndFor
\State $RIP \gets \textit{citation\_count  / num\_papers}$

\State $DCP \gets \textit{Average number of 1-3 year old cited references contained in papers in the dataset citing the target journal}$

\State $median \gets \textit{median DCP of all journals}$

\State $RDCP \gets \textit{DCP  / median}$

\State $SNIP \gets \textit{RIP  / RDCP}$
\State \textbf{return } $SNIP$

\end{algorithmic}

\end{algorithm}

   \par A random set of journals in Computer Science and Mathematics were selected and Algorithm 8 was used to calculate SNIP for these journals. The SNIP values thus obtained for the year 2010 with citation data taken from the Aminer Citation Network Data Set \cite{26} were compared with their actual values for the same journals and same year provided by Journal Metrics \cite{3,4}. The values thus obtained were not on par in terms of sheer magnitude with the corresponding values provided by Journal Metrics due to two main reasons.
Firstly, our database is a fraction of the one used by Journal Metrics in terms of size, and further, some citations in our database may not have been included by Journal Metrics in calculating SNIP - and vice-versa. 
\vspace{2mm}
\par \noindent However, on further analysis using regression, we were able to show that there does indeed exist a strong correlation between the two values - SNIP calculated by us and the actual SNIP values provided by Journal Metrics. Figure 11 shows a linear regression line, fitting calculated SNIP and actual SNIP. We obtain an R-squared value of 0.7363 meaning 73.63\% of the variance in actual SNIP is accounted for by the variance in calculated SNIP.

\begin{figure}[h]
  \centering
  \includegraphics[width=0.4\linewidth, height=6cm]{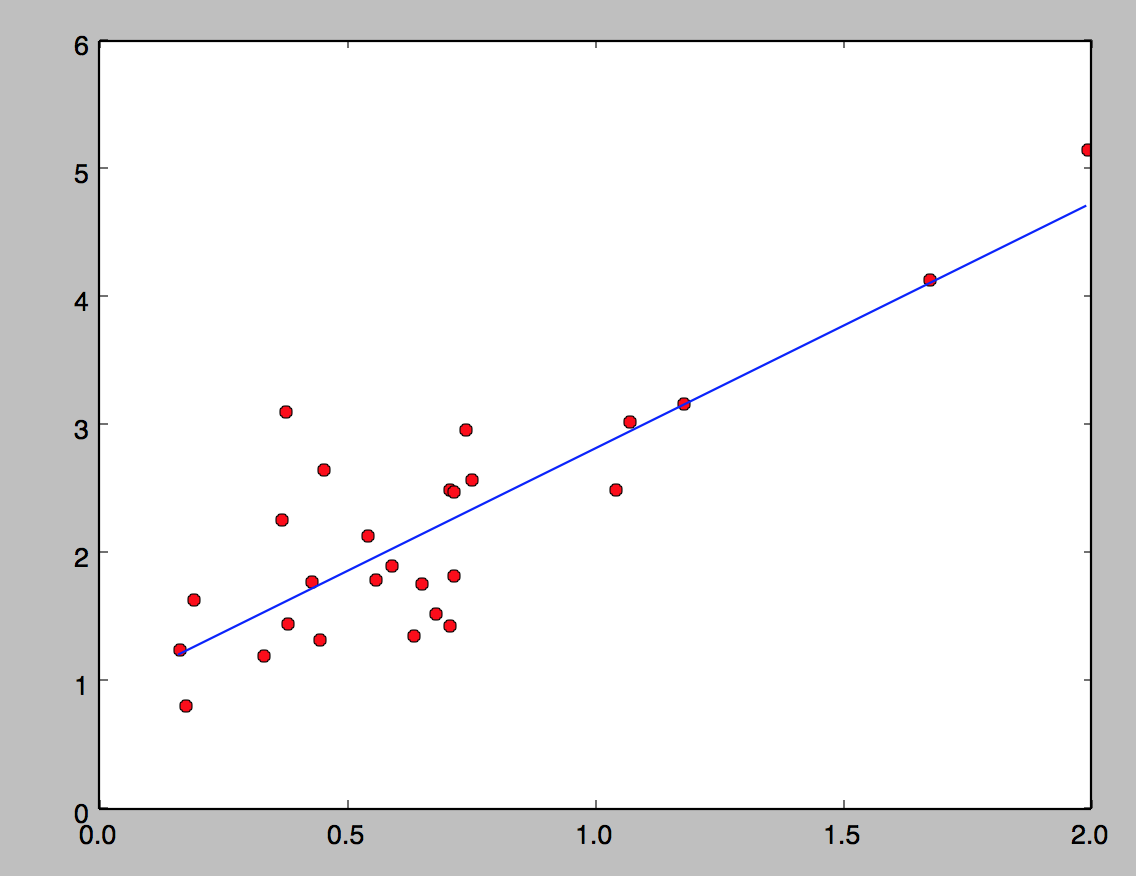} 
\caption{Actual SNIP v/s Computed SNIP}
  \label{fig:subim1}
  
  \end{figure}
           \vspace{2mm}
           
          To further validate our algorithm, we calculated SNIP for 189 journals which were common with both the Aminer data set and Journal Metrics' data set (from Elsevier). In this case, simple linear regression would not suffice; support vector regression (SVR) was used instead \cite{23}. Consider a set of linearly separable points, then the support vectors are those points which are difficult to classify and have a direct influence on the optimal location of the decision boundary. SVR is designed to find an optimal hyperplane which divides the two sets of linearly separable points such that an $\epsilon$-margin from either of these support vectors is obtained. 
\vspace{2mm} 
\par \noindent Initially, an RMSE of 1.162693 was obtained for linear regression and 1.163683 for SVR, without any tuning. Further tuning was performed by changing the  values of $\epsilon$ and cost. The range of $\epsilon$ was narrowed from (0,2) to (0.68,0.72) with the cost parameter narrowed from $2^9$ to $2^2$. As a result, an RMSE of 1.116192 was obtained. Figure 12(a) shows how performance varies with $\epsilon$ and cost, with darker blue areas indicating optimal performance.
        
  \begin{figure}[h]
 \begin{subfigure}{0.5\textwidth}
  \centering
  \includegraphics[width=0.7\linewidth, height=6cm]{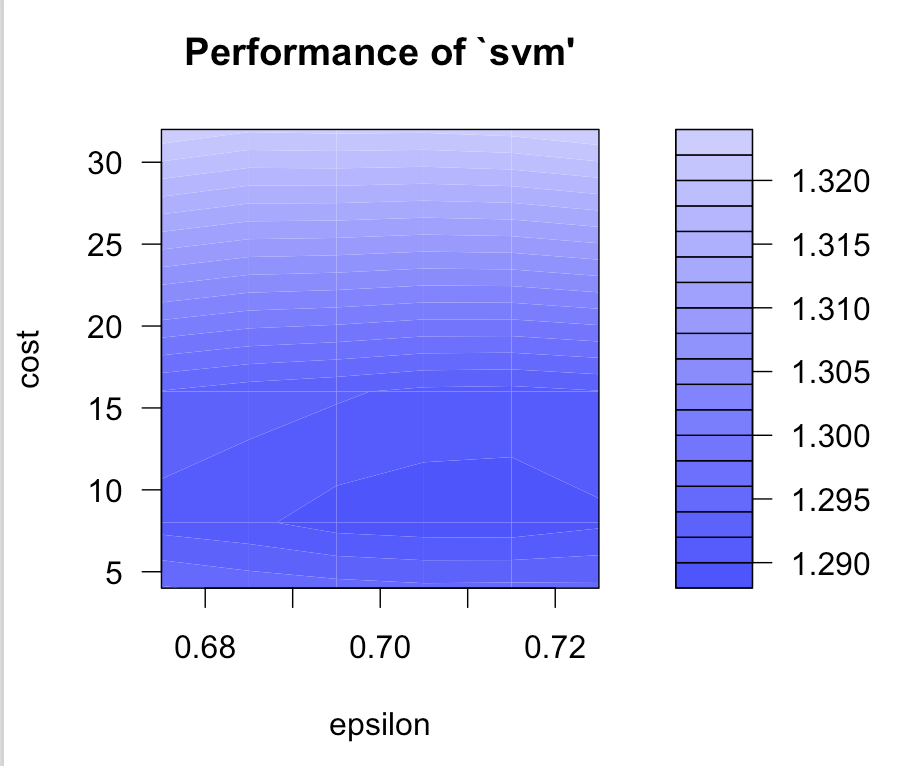} 
\caption{Performance of SVR}
  \label{fig:subim1}
  \end{subfigure}
  \begin{subfigure}{0.5\textwidth}
   \centering
  \includegraphics[width=0.7\linewidth, height=6cm]{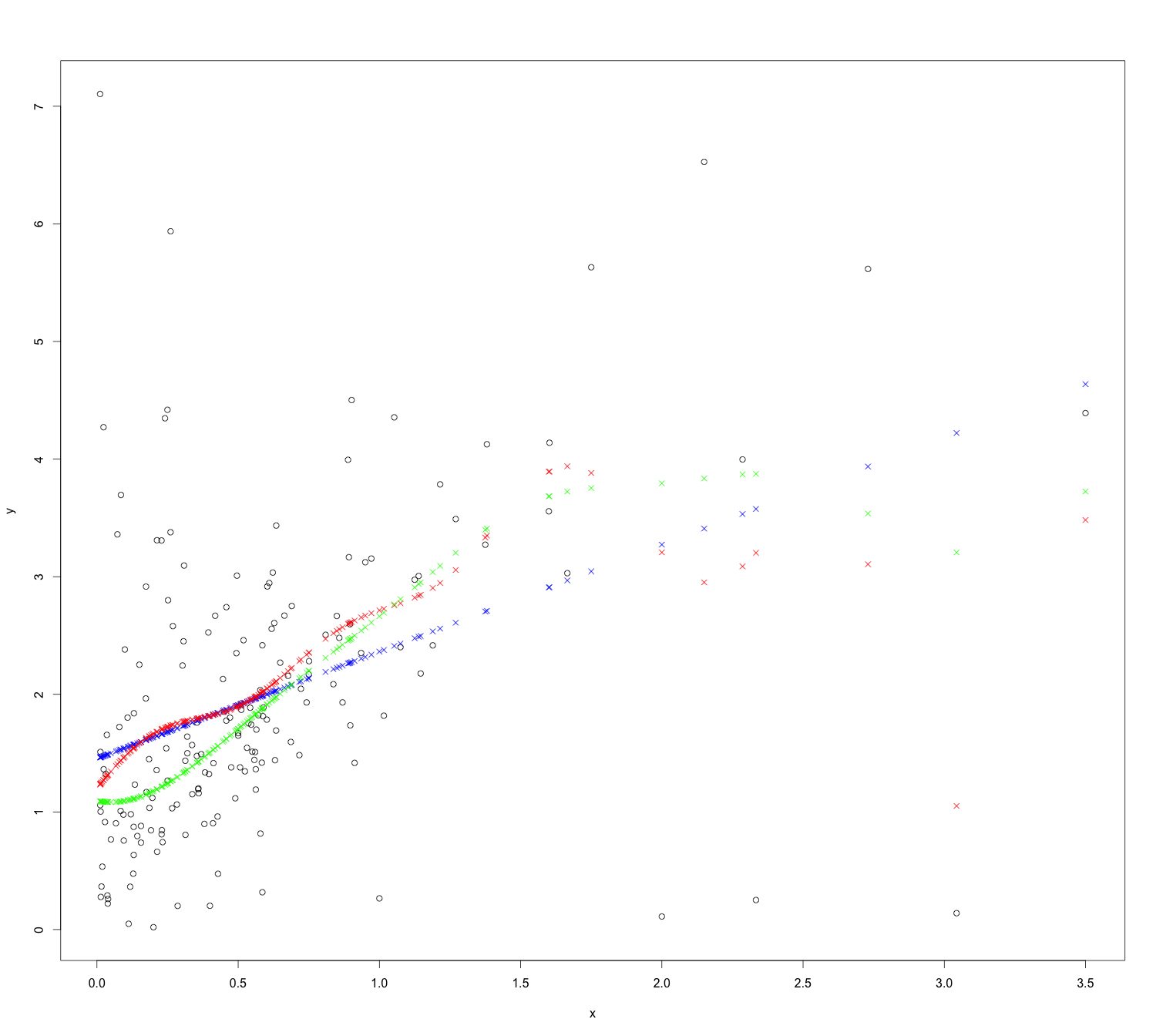} 
\caption{The tuned SVR model}
  \label{fig:subim1}
  \end{subfigure}
  \caption{Results of Support Vector Regression}
\end{figure}

The graph in figure 12(b) shows the linear regression model in blue, untuned SVR model in red and the tuned SVR model in green which gave us the best fit.
  \par To corroborate the results from SVR, exponential and polynomial regression was also performed on the same data set. The exponential model  shown in Fig.13(a) returned an $R^2$ value of 0.9868. The relationship between the Y (aSNIP) and X (cSNIP) along with the coefficients (with 95\% confidence bounds) is best represented by the following equation:
 \begin{equation} Y=1.273 * e^{0.5626*X} \end{equation}
 \par The polynomial regression model shown in Fig.13(b) also resulted in an $R^2$ value of 0.9866.  The equation obtained from polynomial regression is:
 \begin{equation} Y=1.741 * X+0.8641 \end{equation}

 \begin{figure}
 \begin{subfigure}{0.5\textwidth}
  \centering 
  \includegraphics[width=0.7\linewidth, height=6cm]{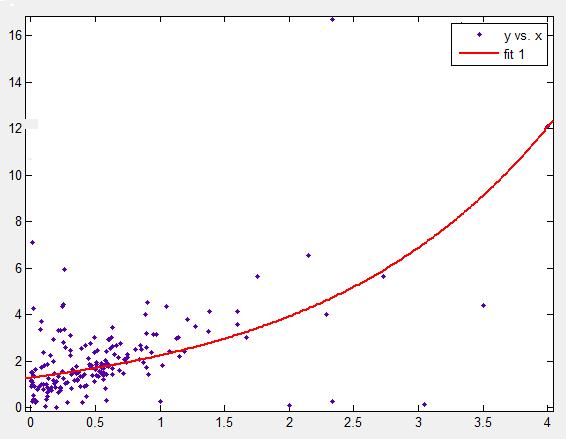} 
\caption{Exponential Regression}
  \label{fig:subim1}
  \end{subfigure}
  \begin{subfigure}{0.5\textwidth}
   \centering
  \includegraphics[width=0.7\linewidth, height=6cm]{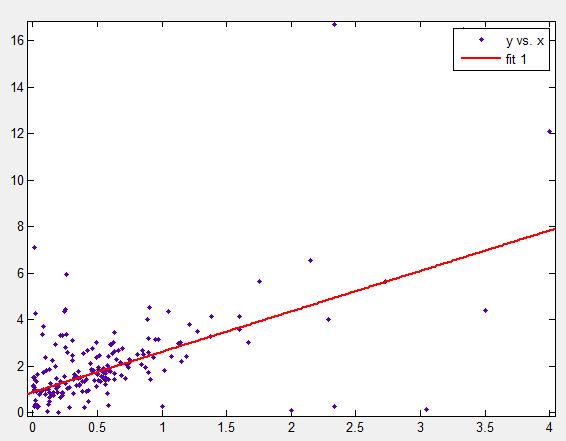} 
\caption{Polynomial Regression}
  \label{fig:subim1}
  \end{subfigure}
  \caption{Curve fit of Exponential and Polynomial Regression}
\end{figure}
 
 We can confidently conclude - with the help of the derived RMSE and R-squared values, graphs and equations - that support vector regression, polynomial regression and exponential regression are all suitable for predicting actual SNIP values from their calculated values, with a very high degree of certainty.

         \vspace{1mm}
         \subsection{Journal Collaboration}
         
         \subsubsection{Inter-Journal Collaboration}
               \par The same journals used in section 7.2 Fig. 11 are considered in this section as well. The citation data was taken from Aminer Citation Network Data Set \cite{26} and correlated with journals having SNIP values taken from Journal Metrics \cite{3,4}. The citation network was then constructed as follows - the nodes are taken as journals, the size of each node is relative to it’s SNIP value taken from Scopus. The edge between two journals is a citation between a paper in one journal to a 
               paper in another journal, where the color gradient of the edge is relative to the number of 
               citations in total using Algorithm 9, as shown below in Fig. 14. 
                \par Figure 15 shows the citation network when we separate the edges on the basis of the table is given in Fig. 16, namely, those originating from and ending at a journal having a SNIP value above the median, and the same for below the median, as well as those going from a journal having a SNIP value above the median to below and vice versa.
                
                \begin{algorithm}
                \caption{$InterJ\_Collaboration$ : Algorithm to show Inter-Journal Collaboration graph}\label{euclid}
                \begin{algorithmic}[1]
                \State \textbf{Input: }$databse \enspace of \enspace citations$
                \State \textbf{Output: }$graph \enspace and \enspace adjacency \enspace matrix \enspace of \enspace inter-journal \enspace collaboration$
                
                \State $data \gets \textit{read aminer\_cites}$
                \State $journals\_low,journals\_high \gets \Call{SplitByMedian}{$data$}$
                
                \State $G \gets \textit{DiGraph}$
                
                \ForAll{publication in data}
                \State $papers \gets \textit{data[publication]}$
                
                \ForAll{paper in papers}
                \State $cites \gets \textit{data[publication][paper]}$
                
                \ForAll{cite in cites}
                \State $src \gets \textit{publication}$
                \State $dest \gets \textit{cite['publication']}$
                
                \If{src = dest}
                \Comment{self cite within publication}
                \State continue
                \EndIf
                
                \If{src in journals\_low and dest in journals\_low}
                \State $type \gets \textit{1}$
                \EndIf
                \If{src in journals\_low and dest in journals\_high}
                \State $type \gets \textit{2}$
                \EndIf
                \If{src in journals\_high and dest in journals\_low}
                \State $type \gets \textit{3}$
                \EndIf
                \If{src in journals\_high and dest in journals\_high}
                \State $type \gets \textit{4}$
                \EndIf
                
                \ForAll{type $\leftarrow$ (1, 2, 3, 4)}
                \If{edge(src, dest) in G } 
                \State $G[src][dest]['weight'] \gets \textit{G[src][dest]['weight'] + 1}$
                \Else 
                \State $G[src][dest]['weight'] \gets \textit{1}$
                \EndIf
                \EndFor
                
                \EndFor
                \EndFor
                \EndFor
                
                \State $G1 \gets \textit{G}$
                \ForAll{edge\_weight in G1 $\leftarrow$ edges}
                \State $edge\_weight \gets \textit{log10(edge\_weight)}$ 
                \Comment{Normalize weight}
                \EndFor
                
                \ForAll{type $\leftarrow$ (1, 2, 3, 4)}
                \State \Call{PlotNodes}{$G1$}
                \State \Call{PlotEdges}{$G1$}
                \State \Call{PlotAdjacencyMatrix}{$G$}
                \EndFor

                  \end{algorithmic}
                  
                  \end{algorithm}
                  
      			\vspace{2mm}
                \par \noindent When we split the citations up into four groups (see Fig. 15, 16) - namely, out of 19,359 inter-journal citations, a very large majority - 57.962\% - are between journals having SNIP values above the median value versus only 5.103\% between journals below the median. The citation network between journals of low SNIP value is quite sparse while that between journals of high SNIP value is dense - this indicates there is far more collaboration among journals of higher prestige or ranking, and little to no collaboration between those journals having a lower SNIP value, despite the fact that half of the journals taken into account were those below the median. This justifies our use of SNIP as a metric for collaboration - higher the SNIP, more the collaboration.

                \begin{figure}
                \centering
                \includegraphics[width=0.7\linewidth, height=6cm]{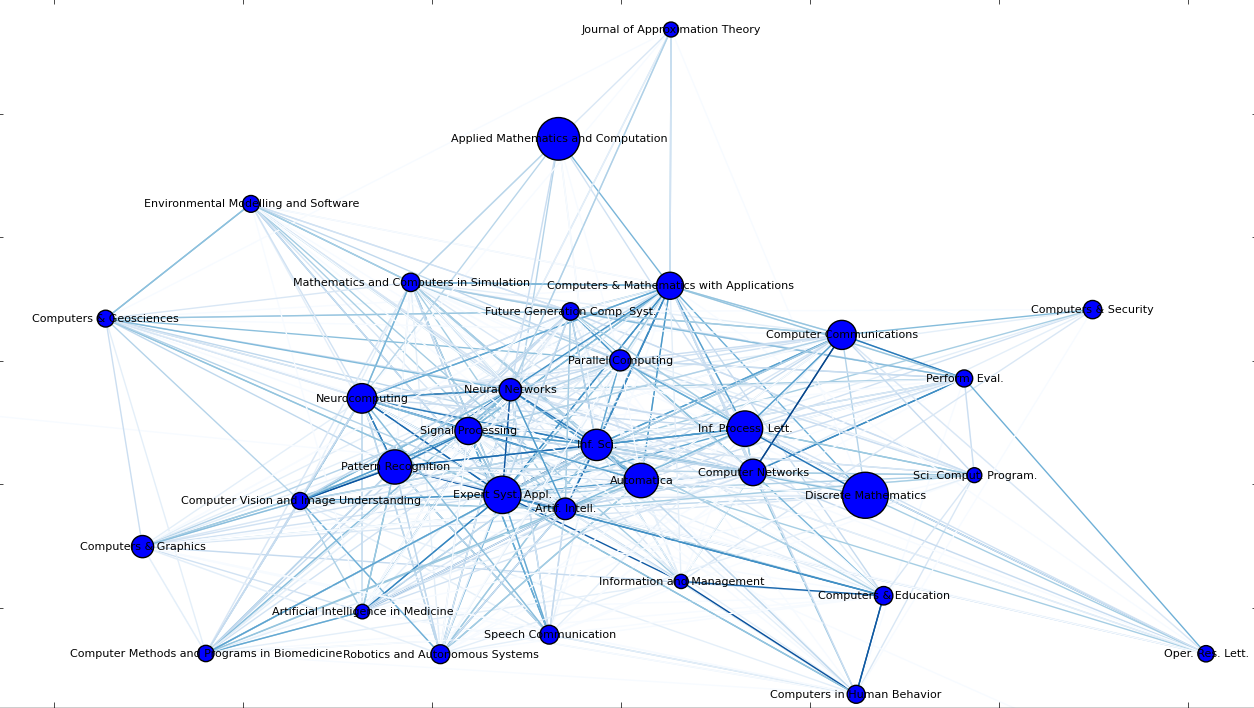}
                \caption{Citation network between all journals}
                \end{figure}
                
                \begin{figure}
                \centering
                \includegraphics[width=11cm]{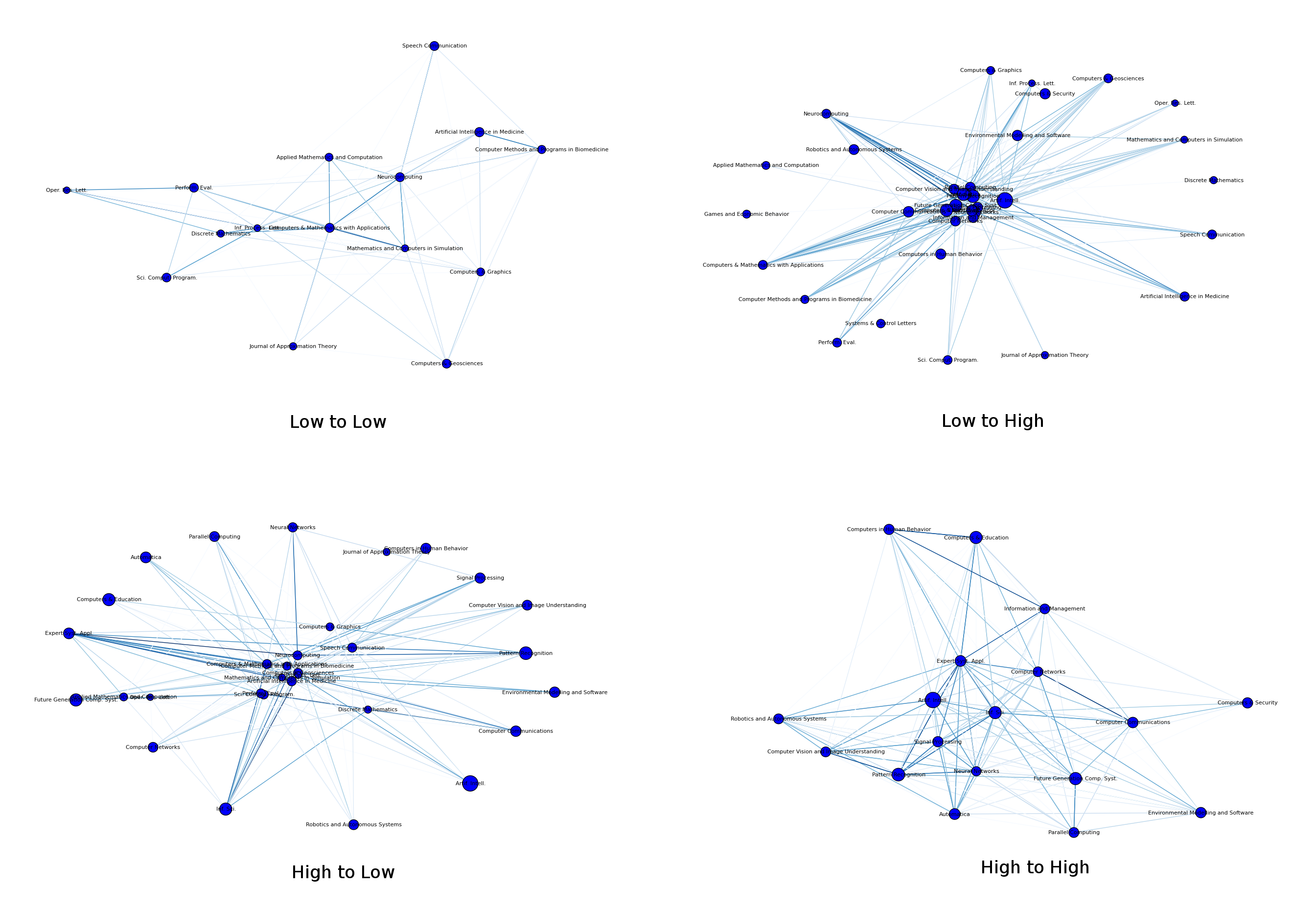}
                \caption{Citation network between journals separated along median SNIP value}
                \end{figure}
                
  \begin{figure}
  \centering
  \includegraphics[width=0.9\linewidth, height=4.5cm]{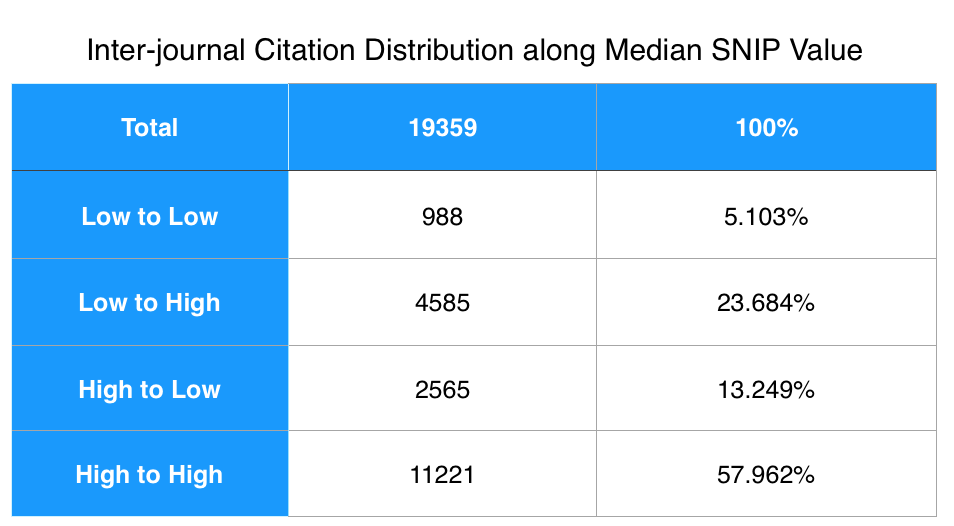} 
\caption{Inter-Journal Citation Distribution}
  \label{fig:subim1}
  \end{figure}
  
           \vspace{2mm}
           \par \noindent One must be careful to note, these numbers could imply that not only do authors tend to favor their papers being published in prestigious journals, but they also cite those papers present in journals of similar level, or papers of the same journal itself. In turn, a cycle is created - authors who publish in prestigious journals are cited more often than those who publish in less prestigious one - thereby increasing the apparent prestige of that journal due to the increased citation count. Whether these citations are genuine or simply reciprocal in nature is not known.
         
          \par \noindent This factor also fuels the growth of predatory journals with nary an oversight in terms of authentic peer-review - less prestigious journals exist with minimal collaboration simply because there was a low bar for a paper to be accepted.

             \begin{figure}[H]
             \centering
             \includegraphics[width=70mm]{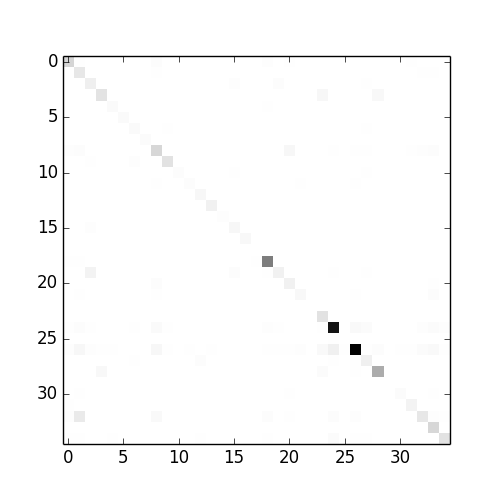}
                  \caption{Adjacency matrix for citations between all journals}
                \end{figure}

               \subsubsection{Intra-Journal Collaboration}
                \par  The adjacency matrix between the journals shown in Fig. 14 and 15 are given in Fig. 17 and 18, with the color gradient being relative to the total number of citations between the journals. It is observed that there exist darker points along the principal diagonal indicating more collaboration within a journal as opposed to inter-journal collaboration. 
                
     \par  Papers published in one journal cite papers from the same journal much more often than those from different journals, regardless of the journal's SNIP value. This, too, leads to a cycle wherein an individual journal's prestige is increased by virtue of increased citations from within. It should be noted that journals of higher SNIP value have a lower NLIQ value as shown in Fig. 19, compared to journals of lower SNIP value - meaning citations are mostly restricted to the same journal they originate from. This in no way implies that there is a correlation between the two (as shown in section 7.4); merely revealing that journals most people would consider to be highly ranked (i.e by having higher SNIP values) exhibit only a low level of non-local influence.
Evidently, information about the internationality of these journals is incomplete - whether the authors are from the same institution or the same country or merely citing their previous works or those of colleagues due to reciprocity, as previously mentioned - is not known.

        \par These are all factors that can be heavily gamed to enhance the prestige and rank of an author as well as the journal their papers are published in. Hence, we  proposed NLIQ in section 5, which favors inter-journal collaboration as opposed to intra-journal collaboration thereby accounting for non-local diffusion of influence and fortifying our definition of internationality. 

\end{enumerate}

                \begin{figure}[H]
                \centering
                \includegraphics[width=11cm]{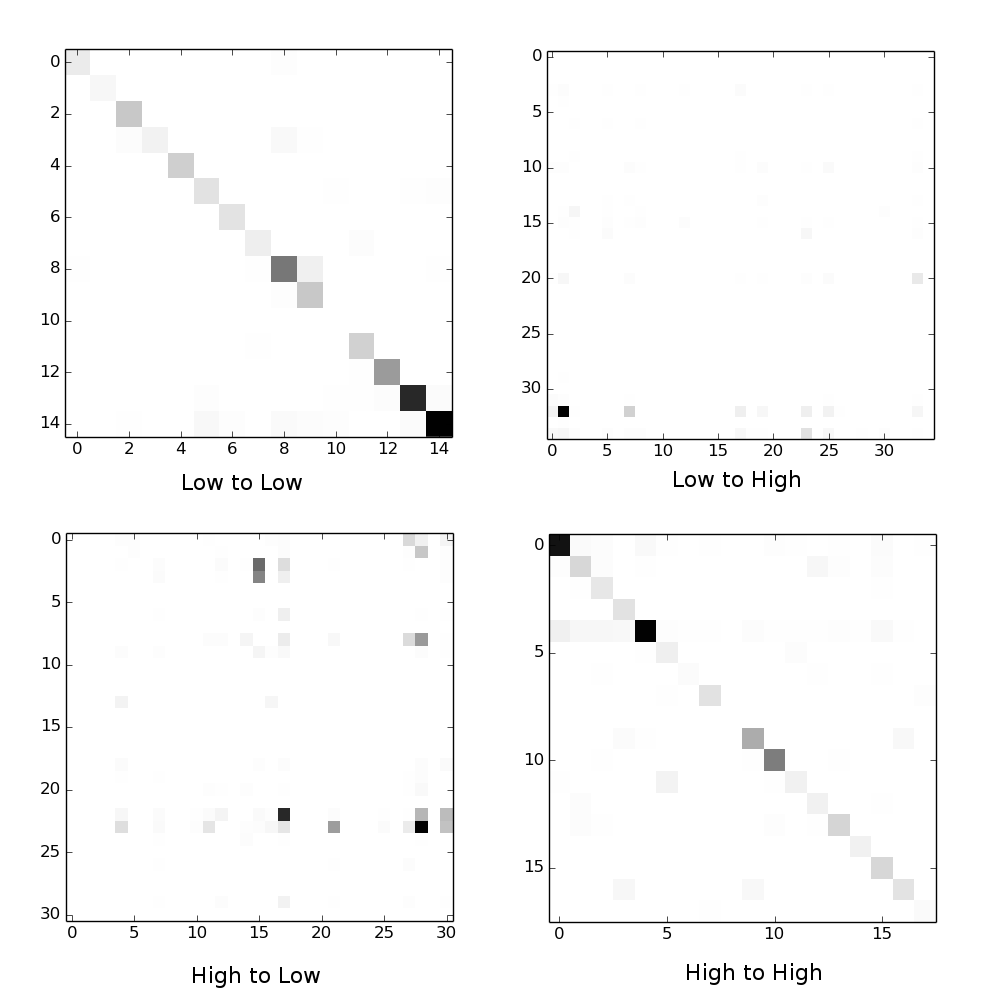}
                \caption{Adjacency matrix for citations between journals separated along median value}
                \end{figure}
              
               \begin{figure}[H]
             \centering
             \includegraphics[width=11.5cm, height = 4cm]{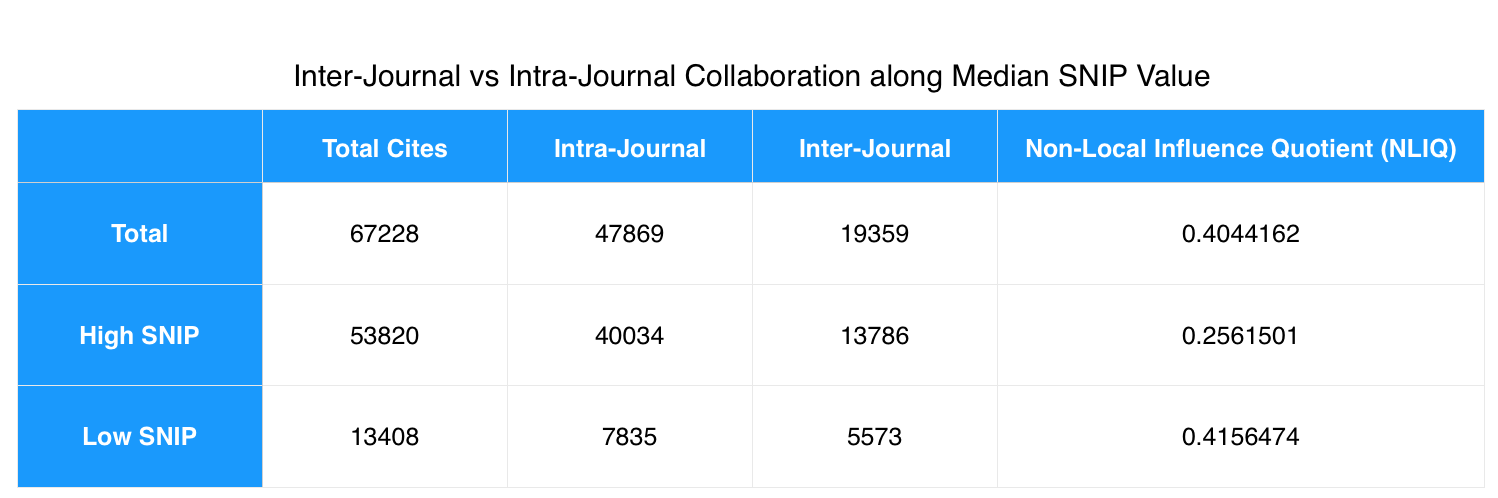}
                  \caption{Non-Local Influence Quotient Statistics}
                \end{figure}
     
     \subsection{Non-Local Influence Quotient, NLIQ}
     
      Reiterating the definition from section 5.3.4, NLIQ is the number of citations made by articles published in a journal X to articles published in different journals divided by the total number of citations made by all papers in that journal X. Clearly, higher the number of external citations made by articles in a journal, higher the NLIQ of that journal.
     \vspace{2mm}
     \par \noindent  
     In Fig. 20, we see a plot of SNIP on X-axis versus NLIQ on the Y-axis. Even though it appears at first that journals with low SNIP values tend to have higher NLIQ, once we look into the goodness of fit and correlation statistics, we see that there is an insignificant relationship between the two. A linear regression line is fit and the R-squared value obtained is 0.1681 and the cross-correlation coefficient as -0.41 at lag 0 and near 0 at lag -1 and +1. A cross correlation value not close to -1 or +1 indicates that there is little to no correlation between SNIP and NLIQ. Similar R-squared values were obtained for higher degree polynomial regression models.
     
     \begin{figure}[H]
             \centering
             \includegraphics[width=65mm]{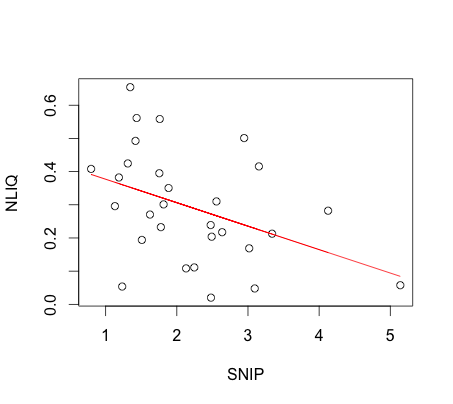}
                  \caption{SNIP vs NLIQ}
                \end{figure}
   \vspace{2mm}  
   \par \noindent 
     Further, even though we don't possess the entire database of citations, we have proven that the Aminer data set has sufficient and even coverage which allowed us to calculate SNIP to a high degree of accuracy - and thus calculating NLIQ with a complete database will not vary by much, either. 
\vspace{2mm}  
   \par \noindent 
Clearly, SNIP is a good indicator of impersonal influence and therefore is used as one of the parameters for computing JIMI, but does not distinguish between the type of collaboration; intra-journal or inter-journal. NLIQ on the other hand is able to differentiate between the two types. It is vital to differentiate between the two; take for example a set of authors who constantly publish papers in one particular journal. They are only collaborating with people in the same area. On the other hand, if papers in different journals and subject areas are cited, cross-collaboration is established. This form of collaboration is indicated by NLIQ, which will be used alongwith SNIP in the Cobb-Douglas model for computing the internationality of a journal.

     \section{Discussion}
There are possibly several parameters which are not considered in the proposed model. We list a few of those which could embellish the metric and internationality score functions.
\begin{enumerate} 
\item Turnaround time(amount of time from the time of submission to publication) and Acceptance Ratio ((Number Of Accepted Papers) / (Number Of Submitted Papers) are two parameters which should be additional measures in computing Internationality Index, However, barring a few journals, this information is not available for scraping. As a future endeavor, we have created a survey to reach out to the editors and hope that this information would be available to us in six to eight months time. Appendix contains the screen shot of the survey. Assuming lukewarm response from the Editor-in-Chief's, this task will have to be accomplished programmatically. The Cobb-Douglas model of scoring is endowed with handling these parameters as long as numerical values could be scraped and computed.
\item Elsevier considers a 3-year window for SNIP mainly due to the difference in the rates at which subject fields mature, whereas Thomson Reuters has a 2-year and a 5-year window for Impact Factor (IF). As noted in section 7.1, one unmistakable advantage of SNIP over IF is that SNIP’s 3-year citation window allows fields that move at a slower pace to be compared with those that advance fairly rapidly, in as fair a manner as possible. Whereas the 2-year IF and 5-year IF only favor one or the other. Thus, authors have considered a window of 3-5 years in order to cater to journals in both categories. Another reason is that many Journals shutdown due to various reasons in a very short span of time. Hence any journal needs minimum incubation time up of 3 years to prove its half-life.
\item An important additional factor, readership profile has to be included as input parameter to the internationality index computation model, which is article, downloads count per country. This could be a challenging task as IP addresses are often masked. The authors don't expect journals to have this feature or co-operate with the authors in order to provide the data, or even allow the authors to fetch the data automatically. Finding a workaround is challenging and the current exercise has no provisions to build on this. This is one weakness that needs to be resolved. 
\item	Currently we consider four predictor variables as input to Cobb Douglas model (Internationality model, JIMI). This model can scale up to any number of inputs, in theory. However, when the numbers of inputs grow in practice the complexity increases exponentially. As the number of input parameters increase, possible curvature violation of the Cobb-Douglas model may create a problem towards estimating the elasticities. However this problem may be resolved by using stochastic frontier analysis \cite{39}. In order to conclude on the exception handling of the model, scale up is needed in future. There are other econometric models more resistant towards curvature violations that the authors intend to explore. From the model perspective, there is a potential curse of dimensionality problem. In that case, use of dimensionality reduction methods become inevitable. These methods help in identifying most significant parameters with high impact on the output when used in the model.
\item	Volumetric information: NLIQ may vary widely across domains and this may hurt some journals more than others. Normalization, not implemented yet in the computation of NLIQ, is a pertinent landmark to accomplish. In order to obtain normalized NLIQ, we could divide the NLIQ of a Journal with the total number of Journals belonging to a domain. The challenge lies in correctly identifying  classification of journals for categorization and count as there is always some overlap across domains. SCOPUS, WOS and GS all have their own logic for segregating the Journals. ACM subject classification is useful and clear enough and might be used for this purpose. The good part of the overlap mentioned above is that journals in niche domains don't get isolated and subsequently NLIQ computation doesn't suffer abruptly. This should alleviate the concern of decent and good Journals having lower NLIQ.
\item	In order to adequately reflect the effect of NLIQ on the internationality score the elasticities need to be adjusted accordingly. The way it should be done includes the choice of an appropriate exponent accompanying NLIQ in Cobb-Douglas Model.
If NLIQ is low but the SNIP is high, we need to choose the elasticities in such a way that it does not hurt the internationality score of the Journal. This is what we call as elasticity boosting, which is achieved through the design of experiment (DoE) study. This study involves computing the percentage contribution of each factor in the Cobb-Douglas model towards internationality score and if contribution of certain factor is low, we adjust elasticity accordingly, at the same time adhering to the constraints of the optimization problem solved during the process. (Theorem 2 of section 6.4 )
\item Estimation of the constant of proportionality, $A$, in the Cobb-Douglas Model: $A$ has been assumed to be $1$ for simplicity in an otherwise complicated computation. However, $A$ in the Cobb-Douglas formulation may be estimated from data by using sophisticated fitting models and constrained optimization techniques. Once $A$ is suitably estimated, elasticities may then be predicted/fitted accordingly.
\item Differentiating citations: Based on the type of article, it is possible to differentiate between number of survey and original research article citations of a journal. Survey or review articles, written tutorials and technical reports tend to receive a large number of citations. It is necessary to distinguish between journals which publish original research articles only, a mix of research and review articles and review articles only(ACM Computing Survey). Therefore, internationality score and parameter quantification need to be normalized accordingly to ensure fair comparison between those journals.
\item Cognizant Citations: It might be a possibility that Editor in Chief's (EiC) are in mutual cognizance and very skillfully suggesting authors to cite articles from journals edited by themselves. For clarity, assume there exists two journals A and B. EiC of A endeavors to boost citation count for B and EiC of B returns the favor. Such cases could be modeled and quantified for penalty in the Cobb Douglas Score function. Graph theoretic modeling might help.
\end{enumerate}
\par It is pertinent to articulate that SNIP alone is not sufficient to compute influence and hence the necessity of defining a metric, NLIQ arises, which disregards the local influence diffusion within a journal. Authors believe that any artificial enhancement of a journal’s influence through coercive citation can be effectively subsided by NLIQ. The paper also uses a parameter - other-citation/total citation which reflects a journal’s integrity owing to the fact that no legitimate journal will promote authors and allow them to indiscreetly cite their own work. Parameters are chosen to ensure that every induced boost to a journals influence through coercive citations, extensive self-citations, copious citations(Definition to follow in next paragraph) or through any other mechanism is negated by the author’s model of internationality and novel metric definitions and computation where ever "conspicuous-citations" are promoted. \\

\par Computing non-local influence and internationality of authors/countries is an important and useful exercise. However, a few questions remain. How wide is the reach of our model? Is the proposed model of internationality portable to authors? If yes, what would be the metrics for such computation and how will the computation be carried out? Is the process consistent with the approach adopted here? Affirmative responses to these questions would strengthen the merits of the model adopted in this paper. As an illustrative example, let us consider NLIQ for authors. The metric definition would vary significantly from the way it is defined for journals. For an author, a local network would include all collaborators, students and supervisor. Therefore, computation of author NLIQ would imply exploring the author's citation networks and genealogy tree. Moreover, it is important to calculate copious citations of authors, if any. "Copious citations" is defined as, if between two authors A and B, say; A cites all published papers of B and vice-versa. This is not difficult to compute but can't be fed directly to the scoring model as other metrics. The reason is straightforward. The proposed scoring model is a constrained growth model and therefore copious citations should be included in the model as penalty subtracted from the main score function each and every time there is an instance of such citation. The model needs to framed as a profit function where revenue function symbolizes the author internationality and penalty is measured as cost function. The current model does not accommodate such cost function definition and penalty estimation.\\
\par Exclusive algorithms meet the requirement for computation of International Collaboration, Other-Citations Quotient and Non-Local Influence Quotient (NLIQ). These algorithms are written to create and develop a platform for \textbf{ScientoBASE: http://pesitsouth.pes.edu/scibase/}, a repository, which will consist of international journals by subject category with ranks and scores of internationality and necessary metric information by using various web-scraping and parsing techniques. Various metrics discuss	ed above are gathered,computed and updated in real-time. This is a major task, once it is ascertained that the proposed model is self-sufficient. It is clear from the discussion and the formulation in previous sections, any new metric or adjustments in the existing metric don't require alterations in the model. However complicated the data assimilation part of the exercise is,  the suite of algorithms help in accomplishing the broader aim of our research in defining a yardstick of scientific contribution and international diffusion; especially in niche areas such as Astroinformatics, Computational Neuroscience, Industrial Mathematics and Data Science from India, as well as other countries across the globe. These are emerging areas and many new journals have come up and for obvious reasons, the metrics are not reported in Scopus and ISI WoS. The outcome of our research will pave way for data and model validation and construction of a data visualization and web interface tool (ScientoBASE Toolkit), an open source web interface, that will compute the scores and provide visualizations of all essential parameters of internationality, particularly for the journals in emerging areas as mentioned above. It is immensely beneficial from pedagogical and scholastic standpoint to be able to use a web-kit and understand the growth of Indian as well as global Scientometry in state of the art and emerging areas in Science and Technology.

 \section{Conclusion and Future Work}
Internationality has thus been defined and perceived as the degree to which a journal transcends local communities and boundaries, with respect to the quality of publication and influence. The methods illustrated in the paper ensures disposition of any kind of local influence that unreasonably boosts scholarly impact of journals. The paper meticulously defines internationality of peer-reviewed journals as a measure of influence that spreads across boundaries and attempts to capture different and hitherto unperceived aspects of a journal for computing internationality.The current work utilizes parameters like "International Collaboration Ratio" which incorporates participation of authors from different demographic regions. Authors humbly submit that "true" internationality of a scholarly publication is necessarily contextual and must be devoid of local or community influence. Source-Normalized Impact per Paper (SNIP) is another parameter taken into account which normalizes the citation pattern within a subject field allowing the comparison of journals belonging to two different domains. This is one of the reasons why authors preferred SNIP over Thomson Reuters Impact Factor (details in section 7.1).\\

\par Exclusive algorithms are written to scrape metric information from the web.  Algorithms to sweep journal, author and article level information needed scrutinization of web pages. However, thanks to the simplicity of Python, it made most of the scraping task hassle-free incurring lesser overhead. Acquisition of journal names, origin, article names, author names and their affiliations were the first few steps in tailoring the parameters of internationality index. These attributes don't contribute to the model directly but are pivotal to creating the indigenous database for subsequent computation. Data acquisition and build are therefore significant contributions of the paper and should not be overlooked.\\
\par The acquired data was validated using different regression techniques. It was observed that values obtained from SNIP algorithm were not very close to the original ones. This is because the native acquisition algorithms could not have scraped through all databases due to the prohibitory firewalls built in several of those. Predictive analytic techniques are used to overcome such barriers so that data recorded in our database is reasonably accurate, endowed with appreciable " goodness of fit " statistic.  The derived RMSE and R-squared values from support vector regression (SVR), linear regression, polynomial and exponential regression were found to be satisfactory. This concludes that any of these regression methods can be used to predict original SNIP (for detailed analysis, refer section 7.2). \\
\par Cobb-Douglas Production Function is used for the first time to model internationality of journals. Appropriateness and adequacy of the model is evaluated in section 6. It is shown that a function is strictly concave if the Hessian Matrix of the second order partial derivatives is negative semi-definite (Theorem 1). The property holds true for the production function implying that the function is strictly concave in nature given that certain conditions on elasticity are met. The importance of concavity lies in validating that the maximum value obtained by the production function is actually a global maxima (Theorem 2 proves this) and the search for such maxima via the model is complete once the maxima is found,by simulation and otherwise (Fig. 10). This maxima is then used as an indicator of highest internationality score and the subsequent neighborhood values may define lower international levels for the same journal. This process is iterated for all journals sweeping through the database.  \\
  
\par Painstaking care has been exercised in creating a knowledge base of citation pattern followed by authors when they publish their work. To investigate the pattern, inter-journal collaboration network was created from Aminer and Journal Metric dataset. Dense citation network between journals of high SNIP values validated the fact that authors are not only tempted to publish their work in prestigious journals but are also inclined to cite papers of journals having a higher SNIP value. By doing so, receiving citation in large numbers is assured. The intra-journal collaboration network, on the other hand, is a reflection of author’s tendency to cite the papers published in the same journal, suggesting signs of community behavior practiced within journals. In an attempt to disregard such publishing practice, authors, while computing internationality have considered  parameters that precisely and unambiguously define, measure and render new meaning to the internationality of journals. Thus, Non-Local influence Quotient, \textbf{NLIQ} is a major contribution for computing internationality and could potentially be a metric to be used by peers, the authors believe! \\
  
\par Commensurate with the current work, author's research contributions may be summarized as follows.
  \begin{itemize}
  \item Quantification of `internationality' of peer-reviewed journals as a measure of influence, introducing a novel treatment by defining new parameters and acquiring new data.
  \item Post-acquisition, extensive cleaning is performed and rigorous pre-processing was done to make the data readable and usable for the model.
  \item Definition of Non-Local Influence Quotient (NLIQ): It is determined by computing the ratio of journal's "non-local" citations to its "local" citations. The parameter signifies the spread of a journal's influence outside its boundaries. It is not documented, but common knowledge that external and internal factors are at play to ramp up impact factors of journals, in the form of suggestions to cite articles from the same journal. This explains the importance of Non-Local Influence Quotient \textbf{NLIQ} as it could enunciate the \textbf{bias-corrected} impact of journals by boasting of greater number of non-local citations. Therefore the diffusion of a journal's internationality is not manipulated by local factors if it possesses greater NLIQ i.e closer to \textbf{$1$}. NLIQ is thus, a reasonably trustworthy indicator of internationality and a significant outcome of the manuscript.
 \item Definition of Other-Citations Quotient: If a journal's self-citations/total citations ratio is high, then papers in a given journal more frequently cite other articles in the same journal, than articles in different journals. That is, a high level of intra-journal collaboration is exhibited as opposed to inter-journal collaboration. Hence, journals with a high self-citations/total citations value cannot be rewarded a high internationality score. In fact, such journals must be penalized. Self-citations/total citations needs to be low in order to appreciate influence diffusion. This prompted us to define "Other-Citations Quotient" as $ 1- $ (self-citations/total citations); if self-citations equal total citations for a journal, then a journal's internationality score shall be rendered \textbf{ZERO} since such a trend reflects closed-community behavior and not true internationality, as defined by the authors. 
 \item Definition of "internationality" as a metric that shows evidence of non-local diffusion. To effectively reduce the effect of localization, SNIP is considered as a parameter for influence calculation.
  \item Novel Algorithms: Developing algorithms to compute International Collaboration, Other-Citations Quotient, NLIQ and SNIP as part of research carried out by the authors. The algorithms scrape and compute the required parameters to be fed into Cobb-Douglas model as a part of internationality computation.
  \item Predictive Analytics: Extensive validation process is carried out on the values obtained from scraping algorithms particularly for SNIP. Regression analysis and support vector regression is performed to confirm these values and the results are found to  meet  the expected level. Elaborate simulation and testing support the validity of our results. 
 \item Normalization: The input parameters fed to the Cobb Douglas function for computing internationality score are normalized. For example, NLIQ takes into consideration the ratio of citations (external to total) and not raw numbers. This practice allows for a fair comparison between different subject fields such as Computer Science, the Social Sciences and Mathematics where collaboration and citation trends differ remarkably.
\end{itemize}  

\par It is possible to consider parameters such as number of article downloads per country and average cites per country along with the ones already included in the model. The model and data acquisition methods may be extended further to visualize growth of a subject based on region (cartogram), author (geospatial influence), topic and journal (spatial diffusion temporal invariant model). Further, this may be extrapolated to include normalization of scientific contributions and diffusion of scientometric indices in niche areas. Once the normalization practice is put to place, ranking and clustering of journals based on internationality may be proposed. This is not very difficult but should be done carefully as just like any other metric, there could be unfair disparity in the internationality scores of journals. Normalization, thus, is central to this entire exercise.

\par Authors do realize that there exists a plethora of metrics for ranking and scoring mechanisms. A practical approach would be to propose one, supported by the two powerful models, Multiple Linear Regression (used in JIS) for general influence and Cobb Douglas Model (in JIMI) for international influence. Authors intend to compute a single score, \textbf{RAGIS -Reputation and Global Influence Score}, $y_{ragis}$ , as a convex combination of JIS and JIMI. JIS ( \cite{32}, Appendix II in the repository contains details of JIS) computes influence score for journals that are indexed in SJR, Scopus and Web of Science. Computation of JIMI brings many other journals under it's fold. \textbf{RAGIS} would facilitate clustering of journals as demonstrated by supplementary data provided in the authors repository (see note below). This is set as a future goal. The authors endeavor to pursue this line of reasoning, hoping for proliferation to a comprehensive set of journals and to cater to a much larger audience. \\   
\textbf{{Note}:} Additional file on GitHub \cite{32} contains Matlab source code that generates an audio/video interface file. The file demonstrates frames of 3D plot of Cobb Douglas Production function. The file contains sample snapshots of the proposed toolkit, as well as other source code used in the course of this manuscript.

\section*{Appendix}
\appendix
    \begin{figure}[H]
             \centering
             \includegraphics[width=65mm]{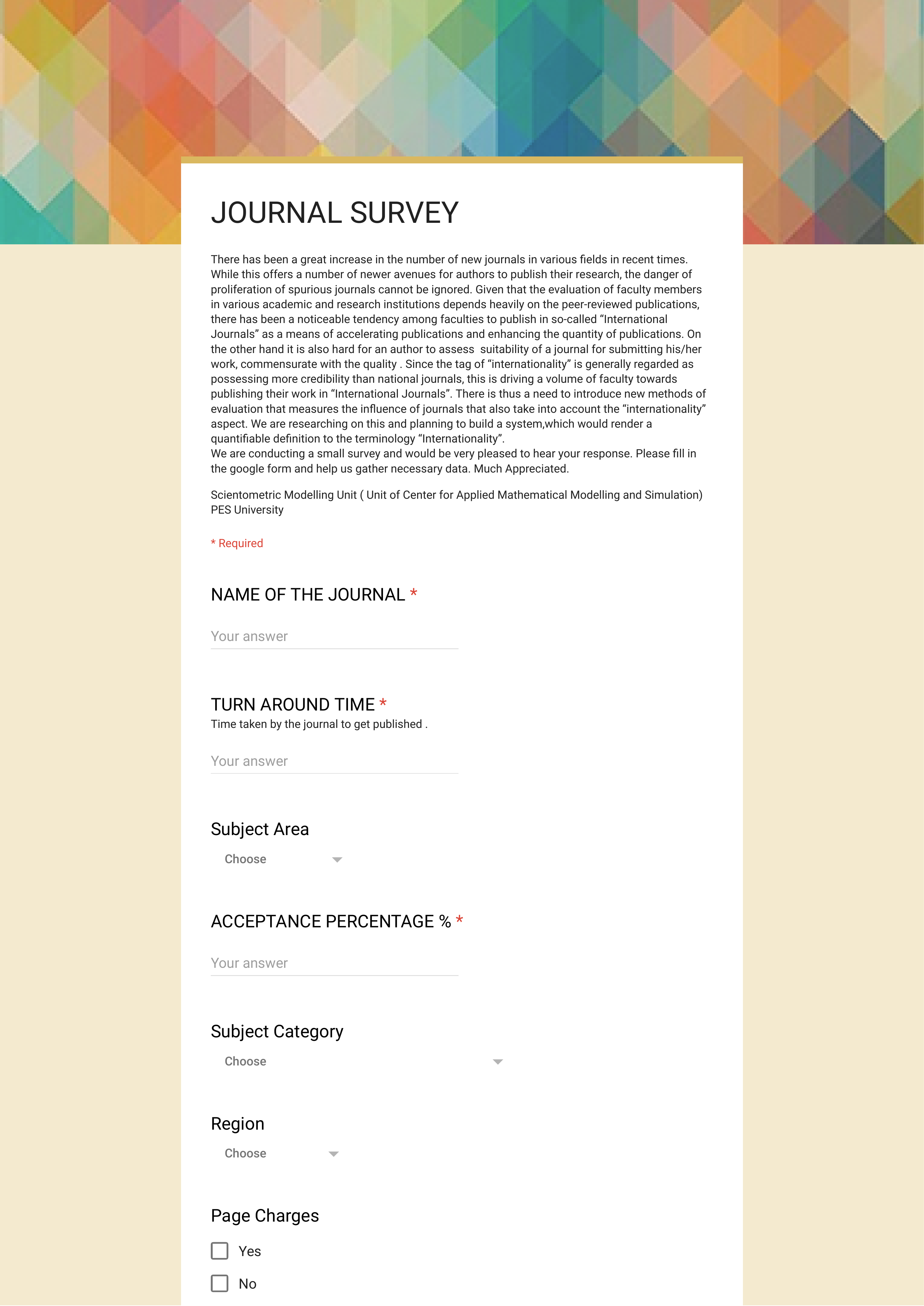}
                  \caption{Screenshot of the survey form}
                \end{figure}
 Link to our website: www.pesitsouth.pes.edu/scibase               
\end{document}